\documentclass{aa} % for a referee version
\usepackage{graphicx}
\usepackage{lscape}
\usepackage{longtable}
\usepackage{arydshln}
\usepackage{multirow}
\usepackage{amsmath}
\usepackage{enumitem}

\usepackage{epstopdf}
\usepackage{natbib}
\usepackage[htt]{hyphenat}
\usepackage[version=4]{mhchem}

\bibpunct{(}{)}{;}{a}{}{,} 
\begin{document}

\title{H$\rm _2$S observations in young stellar disks in Taurus} 
   \author{P. Rivi\`ere-Marichalar\inst{1}, A. Fuente\inst{1}, R. Le Gal\inst{2,3}, A. M. Arabhavi\inst{4,5,6}, S. Cazaux\inst{4,7}, D. Navarro-Almaida\inst{1}, A. Ribas\inst{8}, I. Mendigut\'ia\inst{9}, D. Barrado\inst{9}, B. Montesinos\inst{9}}
 \institute{Observatorio Astron\'omico Nacional (OAN, IGN), Calle Alfonso XII, 3. 28014 Madrid, Spain %1
                   \email{p.riviere@oan.es}
                   \and
   Center for Astrophysics \textbar~Harvard \& Smithsonian, 60 Garden St., Cambridge, MA 02138, USA %2
   \and
   IRAP, Universit\'e de Toulouse, CNRS, UPS, CNES, 31400 Toulouse, France %3
   \and 
   Faculty of Aerospace Engineering, Delft University of Technology, Delft, The Netherlands %4
   \and   
   School of Physics \& Astronomy, University of St. Andrews, North Haugh, St. Andrews KY16 9SS, UK %5
   \and
   Centre for Exoplanet Science, University of St Andrews. North Haugh, St Andrews, KY16 9SS, UK %6
   \and
   Leiden Observatory, Leiden University, P.O. Box 9513, NL 2300 RA Leiden, The Netherlands %7
   \and
   European Southern Observatory (ESO), Alonso de C\'ordova 3107, Vitacura, Casilla 19001, Santiago de
  Chile, Chile %8
   \and 
   Centro de Astrobiolog\'{\i}a (CSIC-INTA), Departamento de Astrof\'{\i}sica, ESA-ESAC Campus, PO Box 78, 28691 Villanueva de la  Ca\~nada, Madrid, Spain %9
   }
   \authorrunning{Rivi\`ere-Marichalar et al.}
   \date{}

 \abstract 
%context heading (optional)
{Studying gas chemistry in protoplanetary disks is key to understanding the process of planet formation. Sulfur chemistry in particular is poorly understood in interstellar environments, and the location of  the main reservoirs remains unknown. Protoplanetary disks in Taurus are ideal targets for studying the evolution of the composition of planet forming systems.}
% aims heading (mandatory)
{We aim to elucidate the chemical origin of sulfur-bearing molecular emission in protoplanetary disks,  with a special focus on H$\rm_2$S emission, and to identify candidate species that could become the main molecular sulfur reservoirs in protoplanetary systems.}
% methods heading (mandatory)
{We used IRAM 30m observations of nine gas-rich young stellar objects (YSOs) in Taurus to perform a survey of sulfur-bearing and oxygen-bearing molecular species. In this paper we present our results for the CS 3-2 ($\rm \nu_0 = 146.969~ GHz$), H$\rm _2$CO $\rm 2_{1,1}-1_{1,0}$ ($\rm \nu_0 = 150.498~ GHz$), and  H$\rm _2$S $\rm 1_{1,0}-1_{0,1}$ ($\rm \nu_0 = 168.763~ GHz$) emission lines.}
% results heading (mandatory)
{We detected  H$\rm _2$S  emission in four sources out of the nine observed, significantly increasing the number of detections toward YSOs. We also detected H$\rm_2$CO and CS in six out of the nine. We identify a tentative correlation between   H$\rm _2$S  $\rm 1_{1,0}-1_{0,1}$  and H$\rm _2$CO $\rm 2_{1,1}-1_{1,0}$ as well as a tentative correlation between H$\rm _2$S $\rm 1_{1,0}-1_{0,1}$ and $\rm H_2O$ $\rm 8_{18}-7_{07}$. By assuming local thermodynamical equilibrium, we computed column densities for the sources in the sample, with N(\ce{o-H2S}) values ranging between 2.6$\rm \times$10$\rm ^{12}~cm^{-2}$ and 1.5$\rm \times 10^{13}~cm^{-2}$.}
{}

\titlerunning{H$\rm _2$S observations in young stellar disks in Taurus}
\maketitle

\begin{table*}[h!]
\caption{Sample positions and spectral types.}
\label{Tab:Sample}
\centering
\begin{tabular}{lllll}
\hline \hline
Source name & RA2000 & DEC2000  & Spectral type & Disk class \\
-- & hh:mm:ss & dd:mm:ss & -- & --  \\
\hline
AA Tau & 04:34:55.4222 & +24:28:53.038 & K5Ve & II \\
DL Tau & 04:33:39.0766 & +25:20:38.097 & K7Ve & II \\
FS Tau & 04:22:02.1925 & +26:57:30.331 & M3.5e+M0e & II \\
GV Tau & 04:29:23.7314 & +24:33:00.216 & K7 & I \\
HL Tau & 04:31:38.4719 & +18:13:58.085 & K5 & I \\
RY Tau & 04:21:57.4132 & +28:26:35.533 & K1IV/Ve & II \\
T Tau & 04:21:59.4323 & +19:32:06.439 & K0IV/Ve & I/II \\
UY Aur & 04:51:47.3900 & +30:47:13.552 & M0e+M2.5e & II \\
XZ Tau & 04:31:40.0868 & +18:13:56.642 & M2e+M2e & II \\
\hline
\end{tabular}
\tablefoot{Disk class from \cite{Luhman2010}.}
\end{table*} 

\begin{figure*}[t!]
\begin{center}
 \includegraphics[width=0.24\textwidth,trim = 0mm 0mm 0mm 0mm,clip]{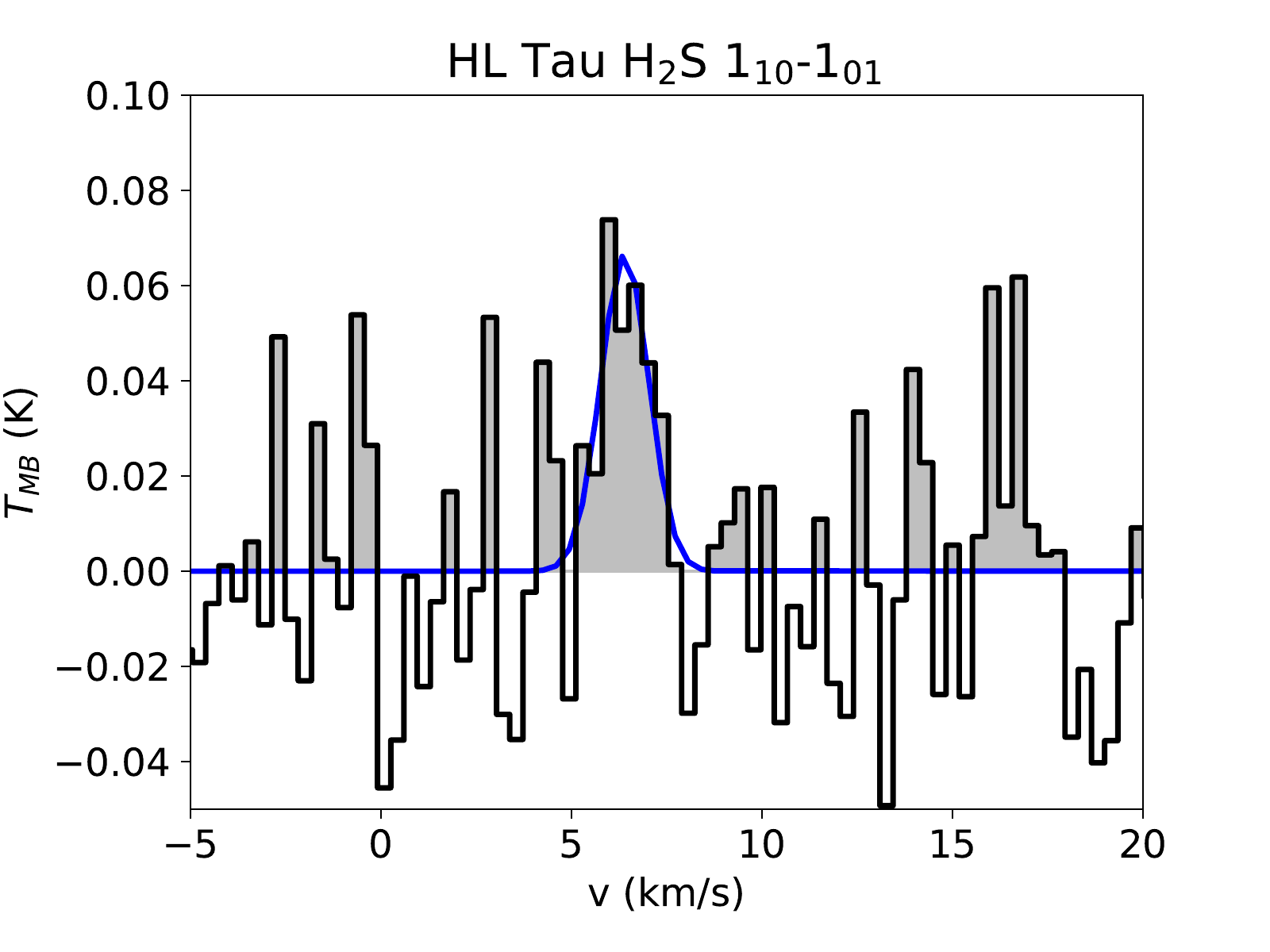}
 \includegraphics[width=0.24\textwidth,trim = 0mm 0mm 0mm 0mm,clip]{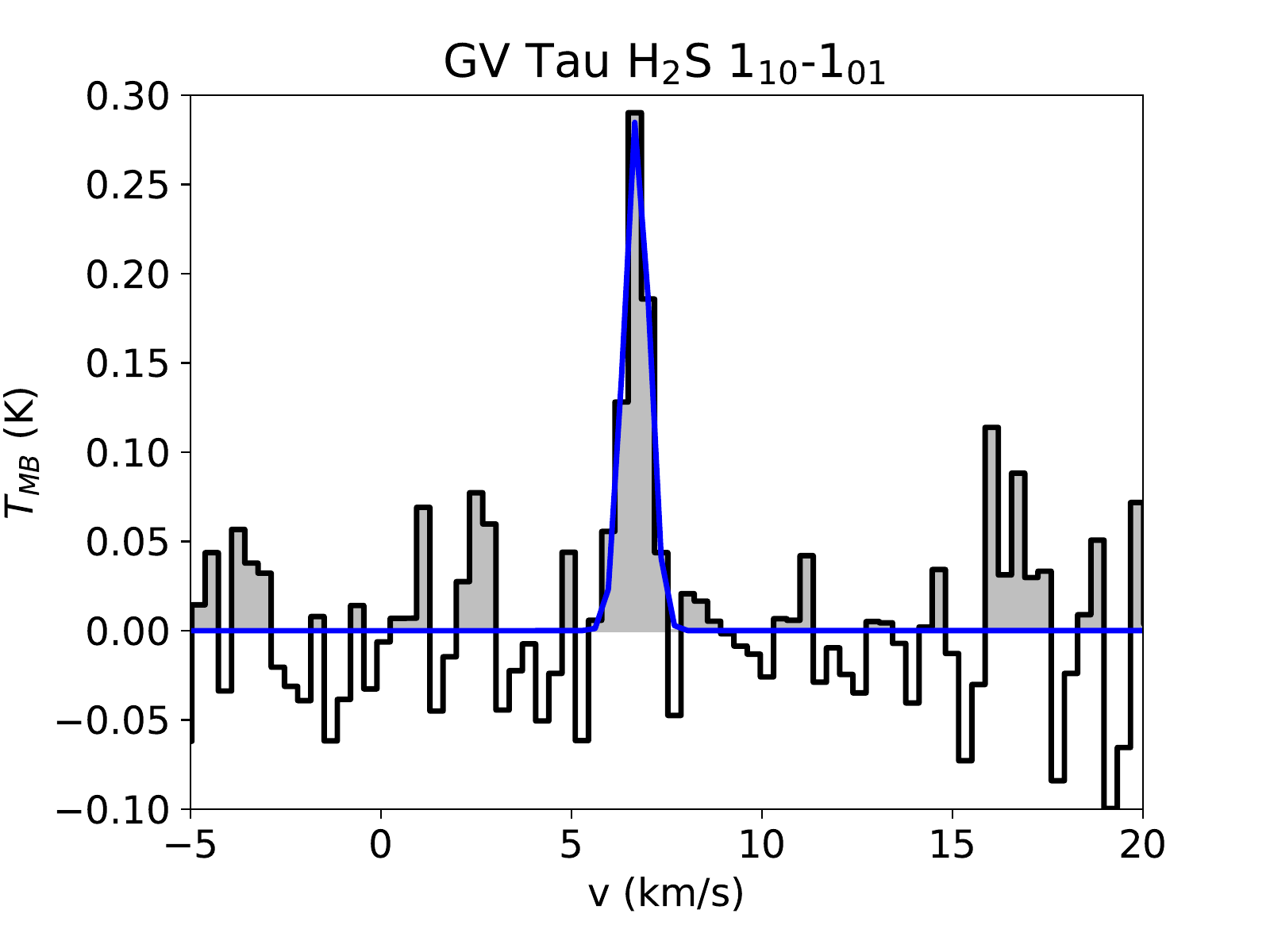}
 \includegraphics[width=0.24\textwidth,trim = 0mm 0mm 0mm 0mm,clip]{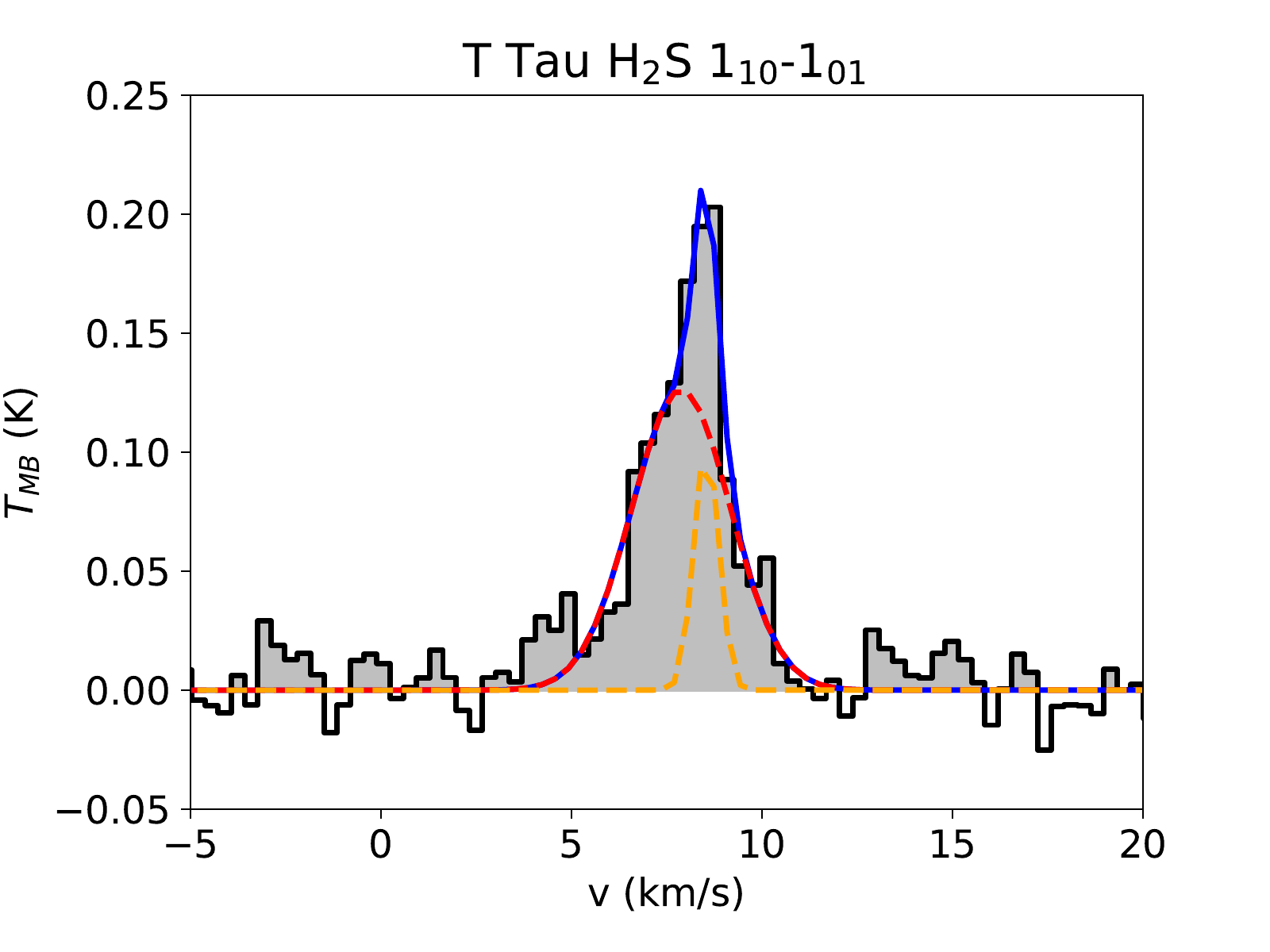}
 \includegraphics[width=0.24\textwidth,trim = 0mm 0mm 0mm 0mm,clip]{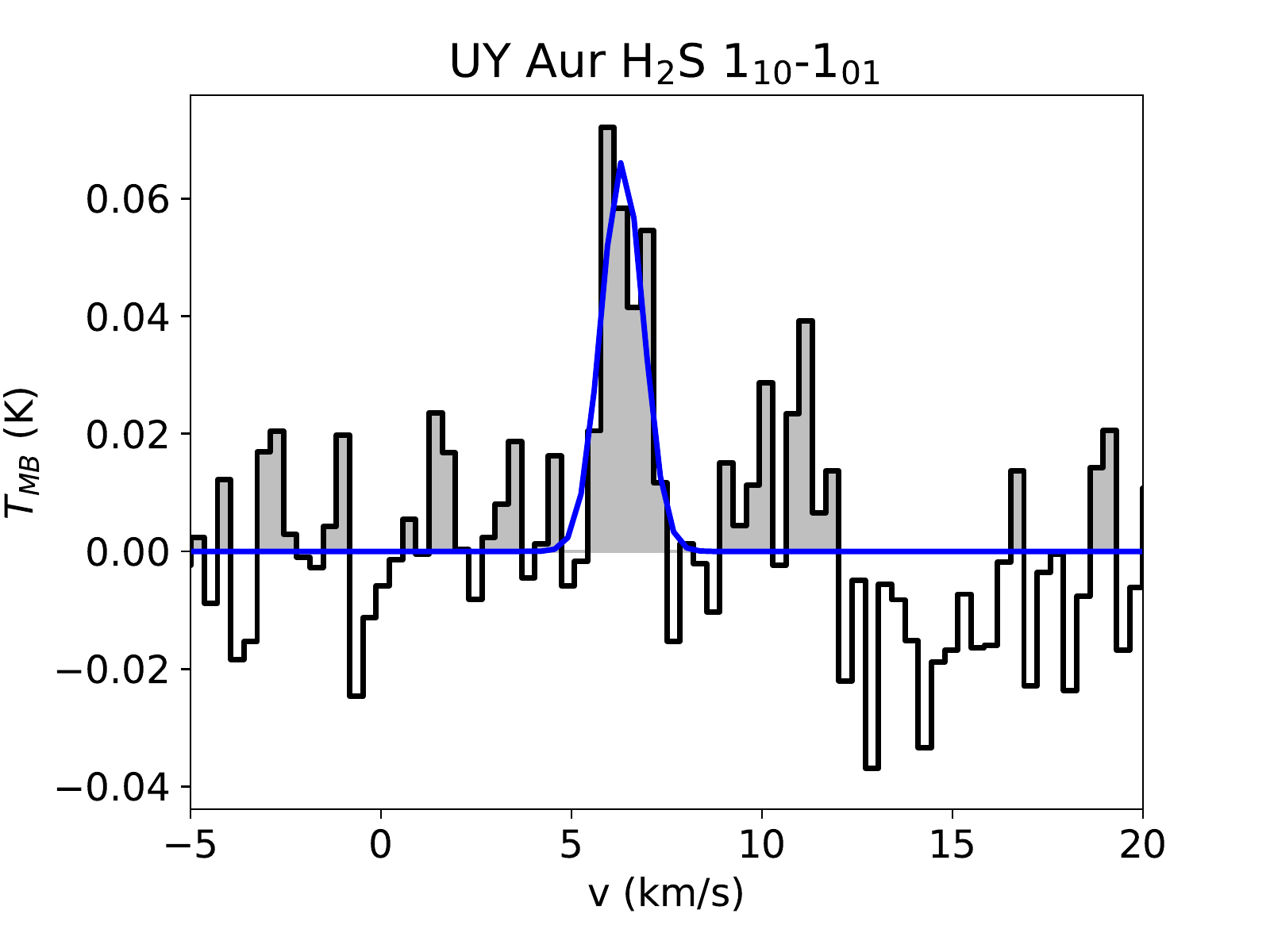}

 \includegraphics[width=0.24\textwidth,trim = 0mm 0mm 0mm 0mm,clip]{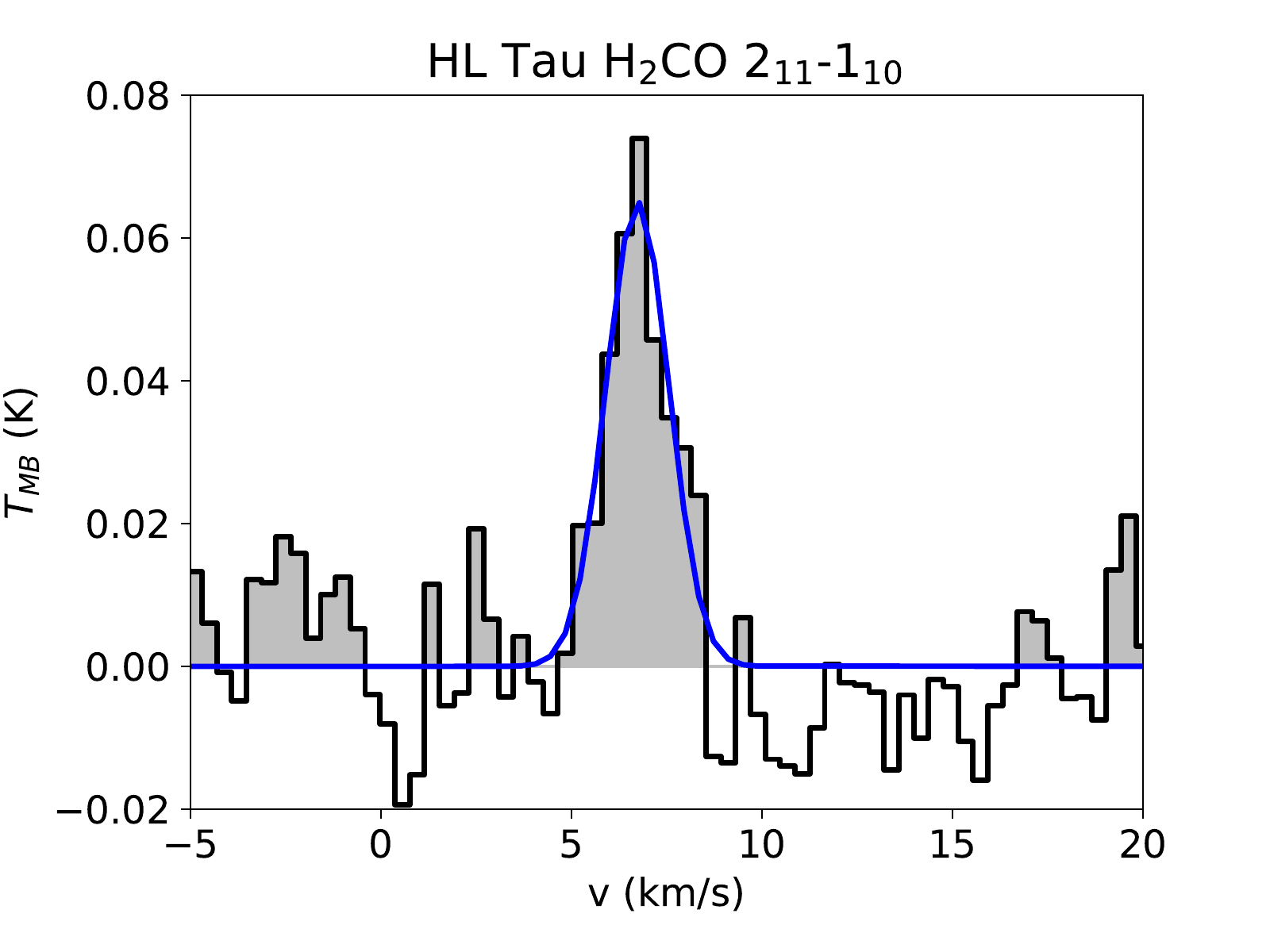}
 \includegraphics[width=0.24\textwidth,trim = 0mm 0mm 0mm 0mm,clip]{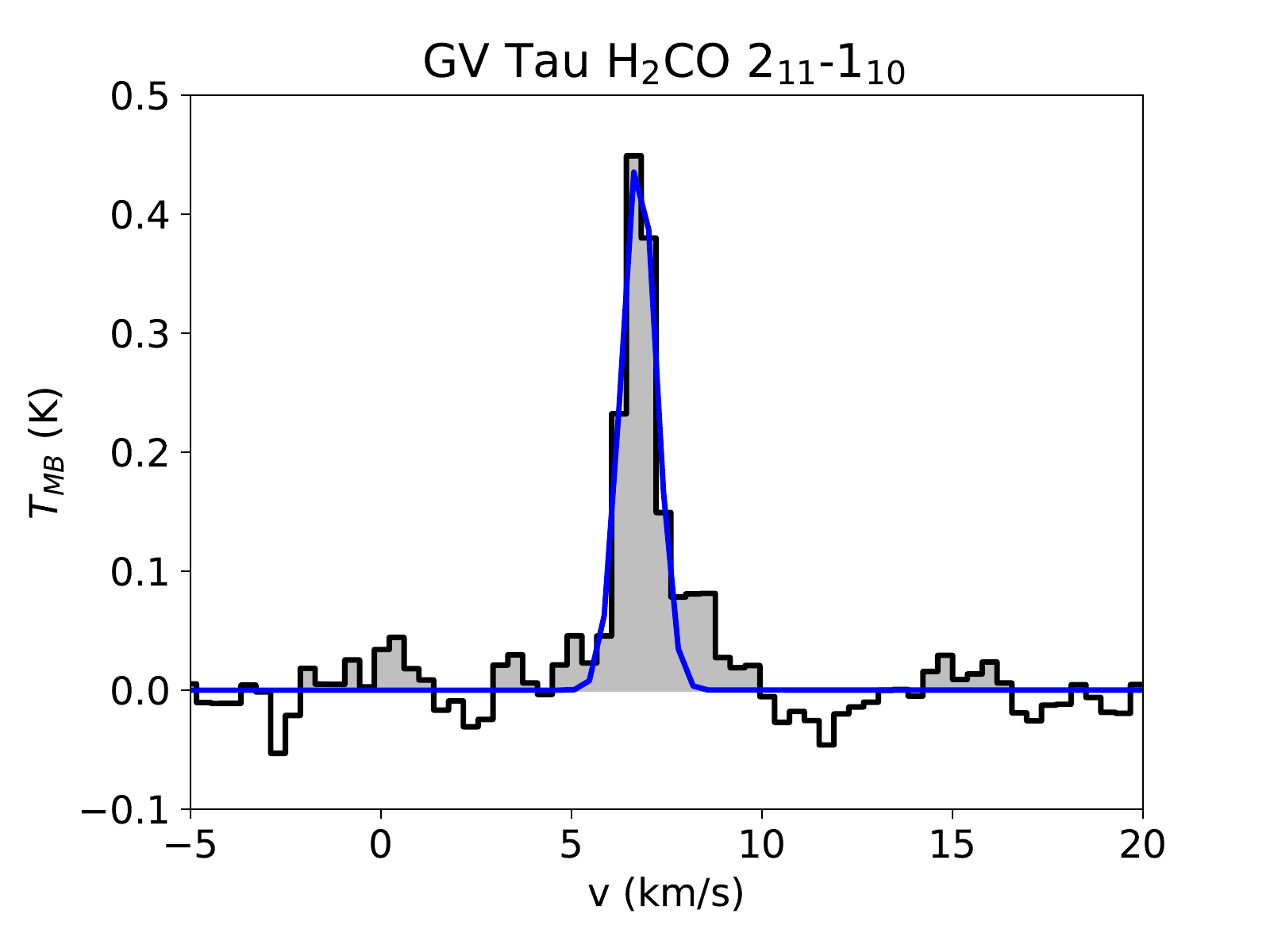}
 \includegraphics[width=0.24\textwidth,trim = 0mm 0mm 0mm 0mm,clip]{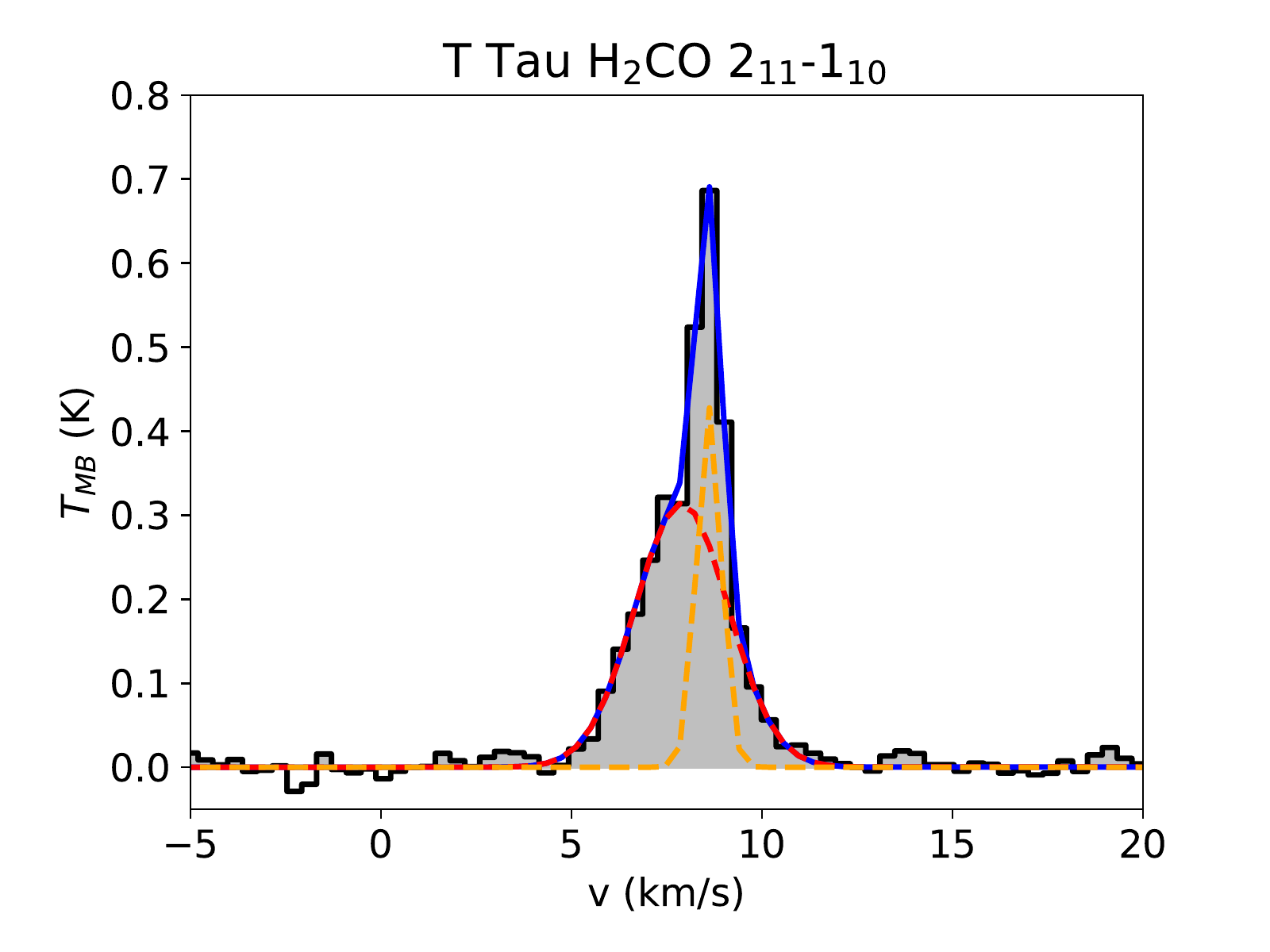}
 \includegraphics[width=0.24\textwidth,trim = 0mm 0mm 0mm 0mm,clip]{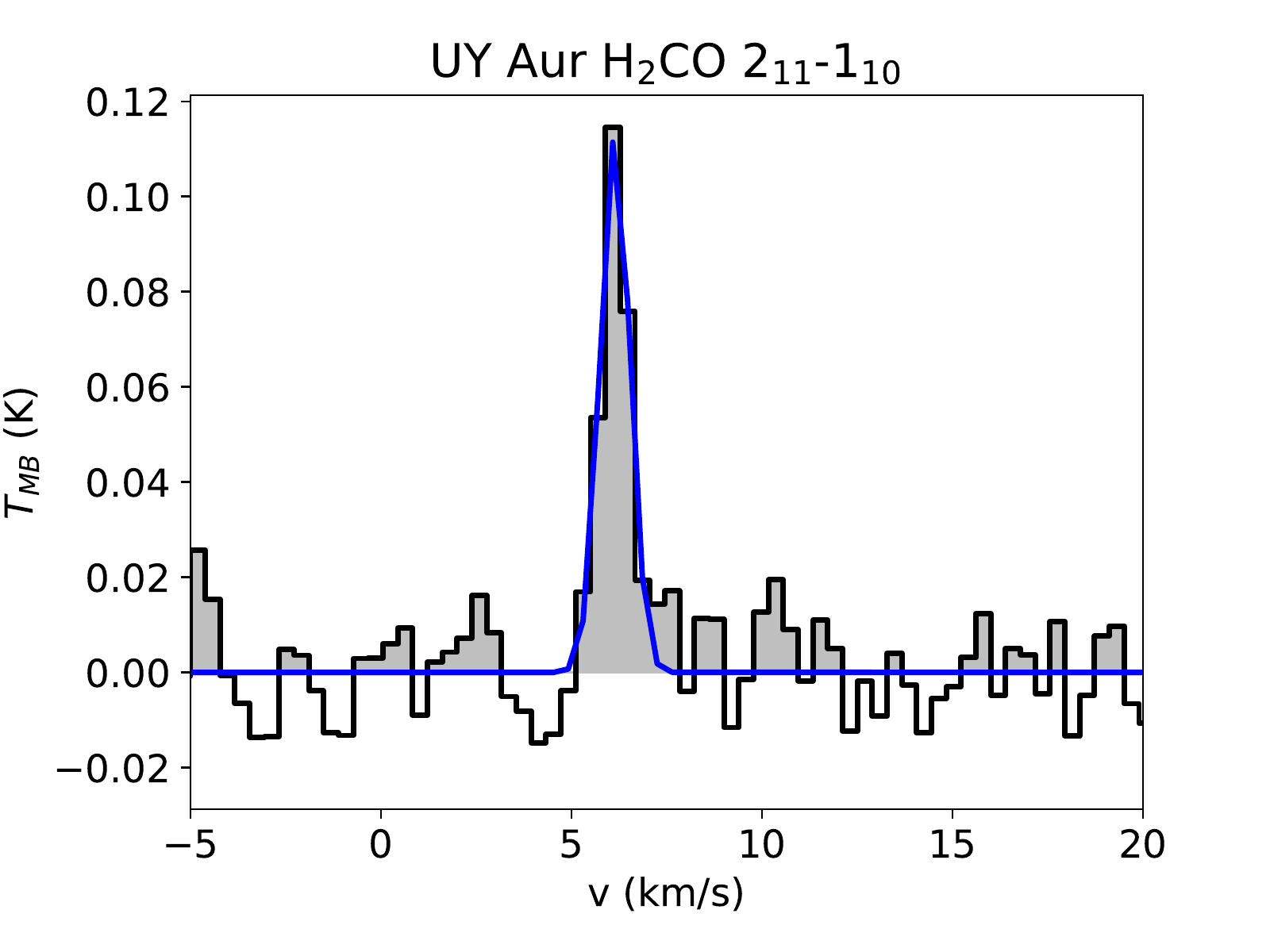}

 \includegraphics[width=0.24\textwidth,trim = 0mm 0mm 0mm 0mm,clip]{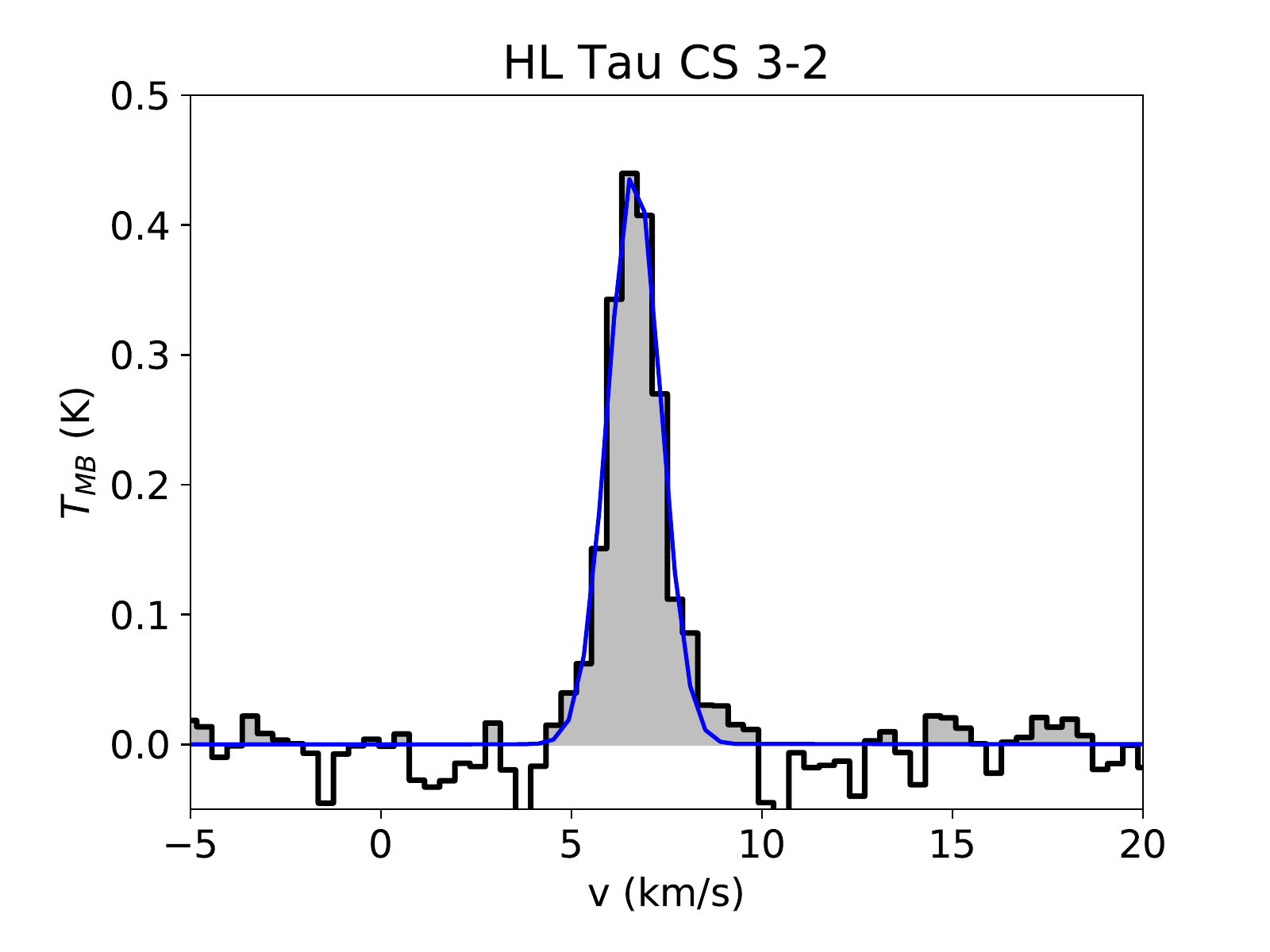} 
 \includegraphics[width=0.24\textwidth,trim = 0mm 0mm 0mm 0mm,clip]{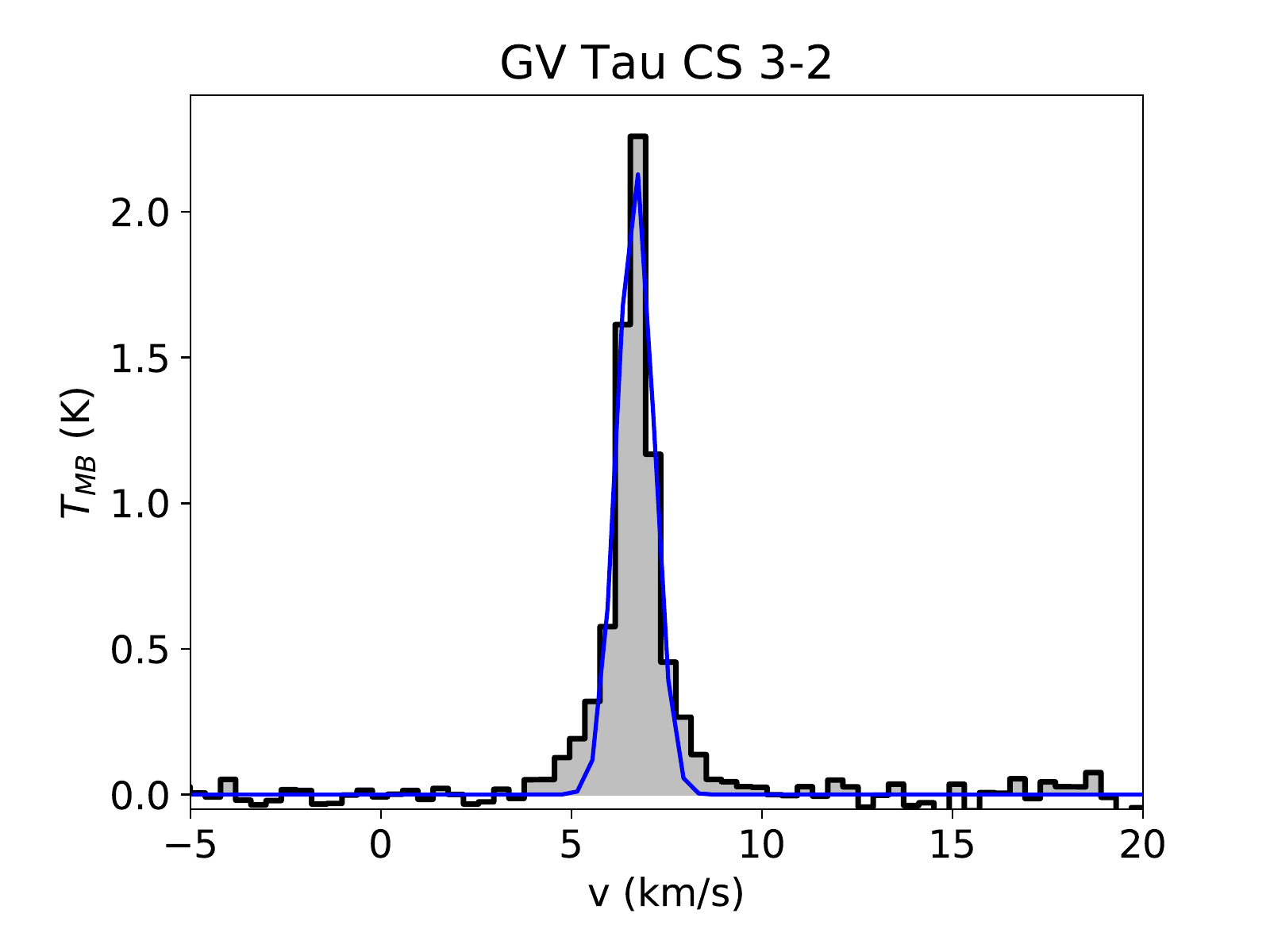} 
 \includegraphics[width=0.24\textwidth,trim = 0mm 0mm 0mm 0mm,clip]{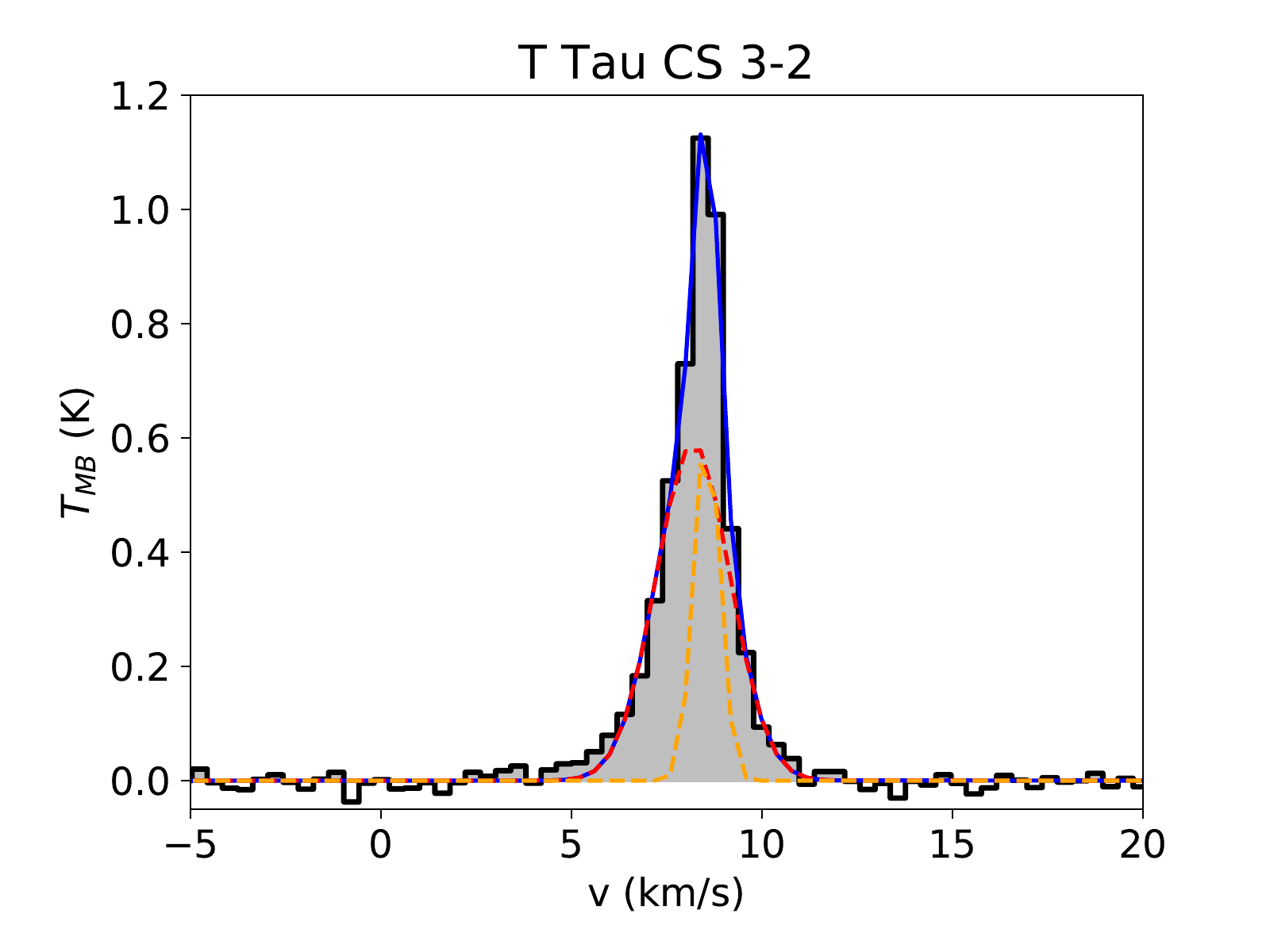}
 \includegraphics[width=0.24\textwidth,trim = 0mm 0mm 0mm 0mm,clip]{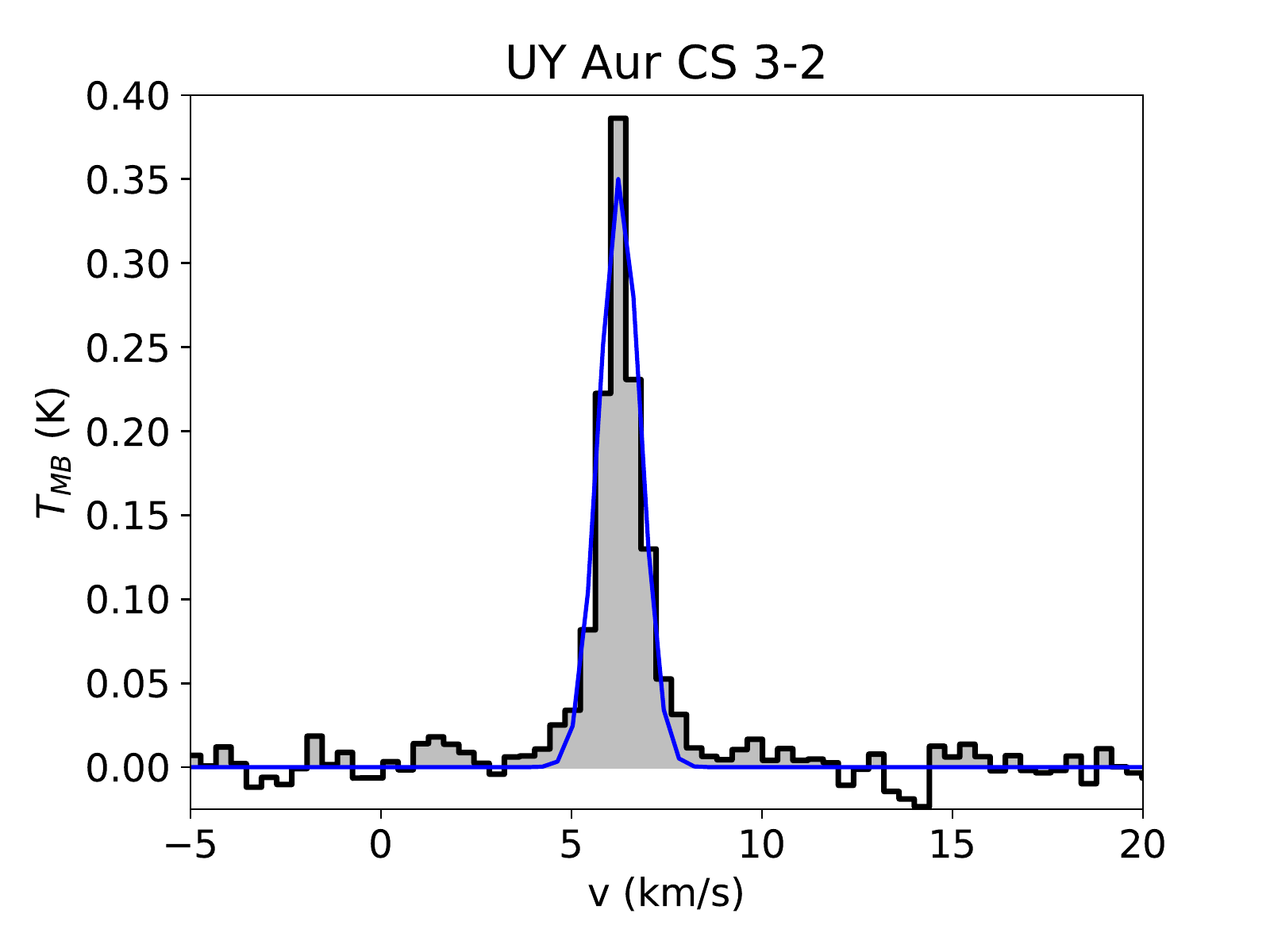} 
  \caption{Spectra of sources with H$\rm _2$S detections. The source name, molecular species, and transition are included at the top of each spectrum. The blue lines represent the Gaussian fit to the observed spectra. In the case T Tau, the blue lines represent a fit with two Gaussians, shown as red and green dashed lines.}
 \label{Fig:spectra_det_H2S}
\end{center}
\end{figure*}

\section{Introduction} 
Planets are born in circumstellar disks that surround young stars. Such disks are made of gas and dust, and they are key to understanding how planets are formed since they fix the initial conditions of the forming planetary systems. Continuum observations have led to a profound understanding of the dust's properties and its spatial distribution. Yet, little is known about the gas chemistry of these systems, even when gas makes up 99\% of the disk mass. Since the discovery of a few molecules more than 20 years ago \citep{Kastner1997, Dutrey1997}, the chemical composition of disks has remained largely unknown. Most of the species detected so far are simple molecules, radicals, and ions, such as CO, $^{13}$CO, C$^{18}$O, CN, CS, $^{13}$CS, C$^{34}$S, C$_2$H, HCN, H$^{13}$CN, HNC, DCN, HCO$^+$, H$^{13}$CO$^+$, DCO$^+$, H$_2$D$^+$, N$_2$H$^+$, c-C$_3$H$_2$, H$_2$CO, H$_2$CS, H$_2$S, H$_2$O, and  HD \citep{Kastner1997,vanDishoeck2003,Thi2004,Qi2008,Guilloteau2006,Pietu2007,Dutrey2007,Phuong2018,LeGal2019a}.  More complex molecules, such as HC$_3$N \citep{Chapillon2012a} and C$_3$H$_2$ \citep{Qi2013b},  have also been detected, but routinely observing them is a challenge; however, CH$_3$CN has been detected in a large number of sources \citep{Oberg2015,Bergner2018}. The molecules that harbor elements linked to organic chemistry (H, C, O, N, S, and P) are of special interest due to their implications for the emergence of life as we know it.

Sulfur is one of the most abundant elements in the Universe \citep[S/H$\sim$1.5$\times$10$^{-5}$,][]{Asplund2009} and plays a crucial role in biological systems. Yet, sulfur chemistry is poorly understood in interstellar environments. It is therefore crucial to follow its chemical history in space and determine which are the main sulfur reservoirs in the different phases of the interstellar medium (ISM).  Sulfuretted molecules are not as abundant as expected in the ISM. In the diffuse ISM and  photon-dominated regions, the observed sulfur abundance is close to the cosmic value \citep{Goicoechea2006,Howk2006}, while in dense molecular gas it is  strongly depleted: Only 0.1\% of the sulfur cosmic abundance is observed in the gas phase \citep{Tieftrunk1994, Wakelam2004,Vastel2018}. It seems that most of the sulfur is locked on the icy grain mantles \citep{Millar1990,Ruffle1999,Vidal2017,Laas2019}. Because of the high abundances of hydrogen and its mobility in the ice matrix, sulfur atoms in ice mantles are expected to form H$_2$S preferentially \citep{Vidal2017}. The only S-bearing molecule unambiguously detected in ice mantles is OCS  \citep{Geballe1985,Palumbo1995};  SO$\rm _2$ has been tentatively detected \citep{Boogert1997}. The detection of H$_2$S, however, is hampered by the strong overlap between the 2558 cm$^{-1}$ band and the methanol bands at 2530 and 2610 cm$^{-1}$. Only upper limits of the solid H$_2$S abundance have been derived thus far \citep{JimenezEscobar2011}. \citet{Laas2019} proposed that other molecules, such as \ce{H2CS}, \ce{CS2}, and \ce{SO}, could be important ice components at later stages of evolution.   Sulfur allotropes, such as S$\rm _8$, were also proposed as possible sulfur reservoirs by \citet{JimenezEscobar2011} and \cite{Shing2020}. The composition of the main sulfur reservoirs remains an open question. 

\begin{figure*}[t!]
\begin{center}
 \includegraphics[width=0.24\textwidth,trim = 0mm 0mm 0mm 0mm,clip]{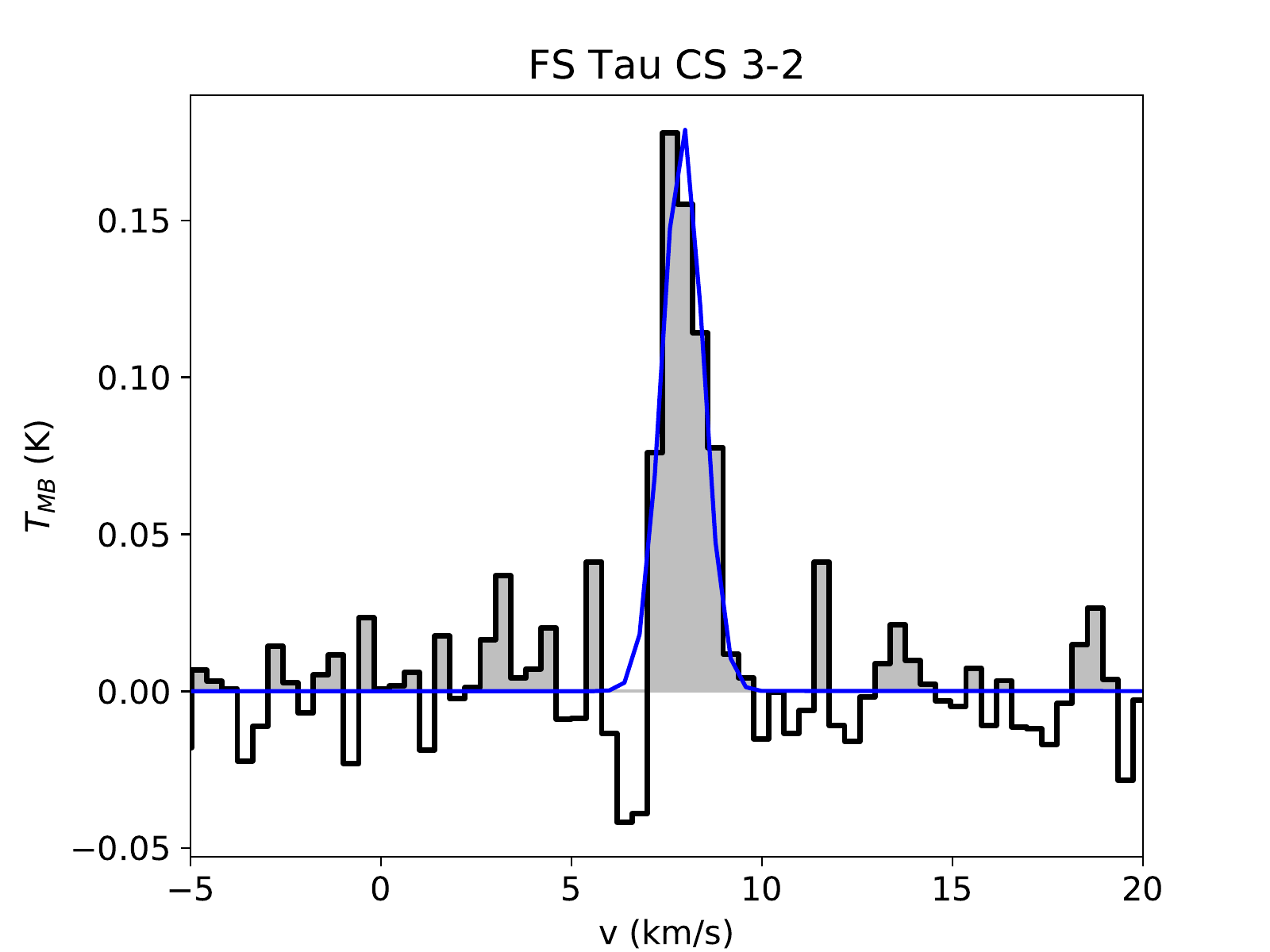}
 \includegraphics[width=0.24\textwidth,trim = 0mm 0mm 0mm 0mm,clip]{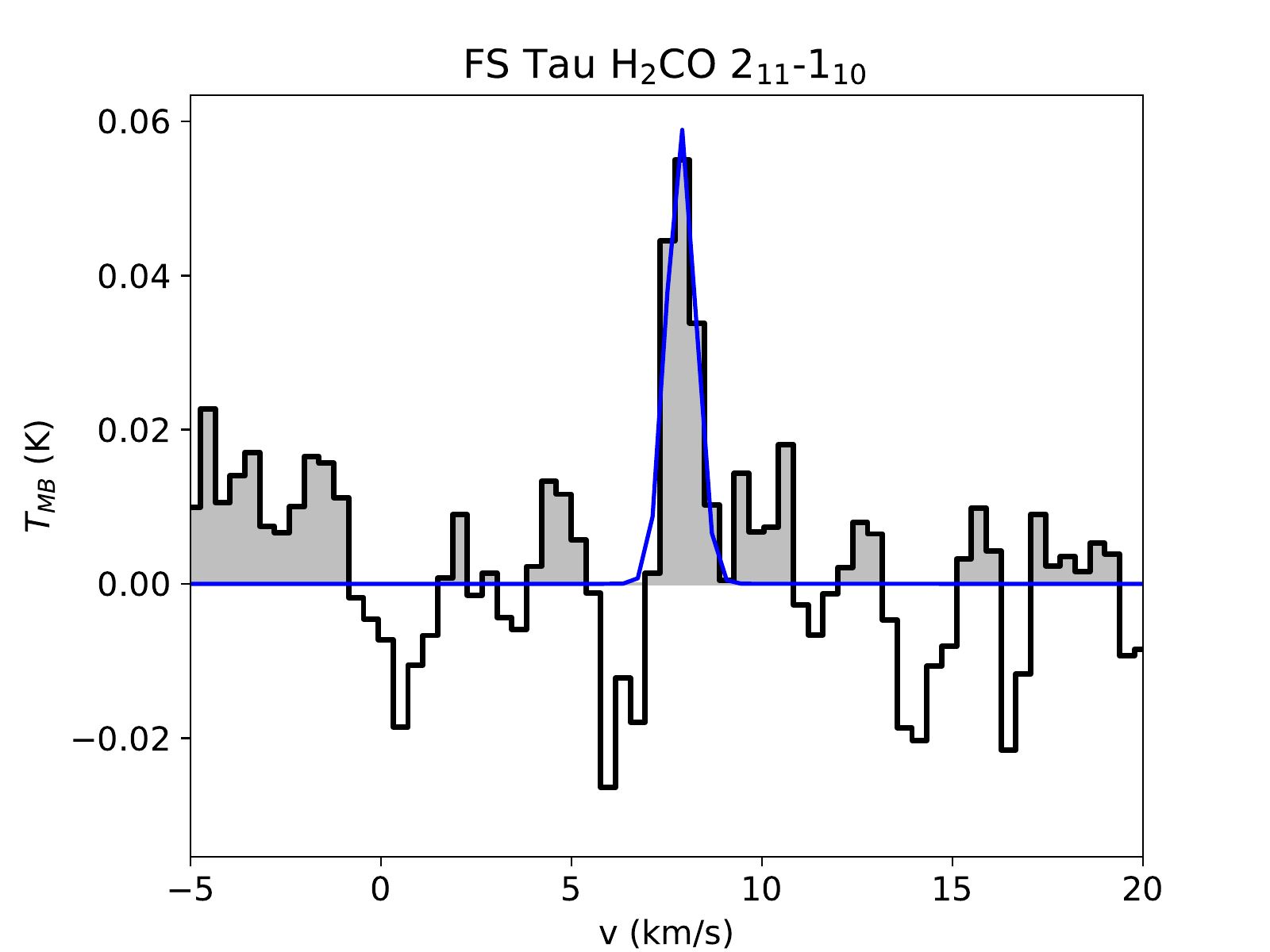}
 \includegraphics[width=0.24\textwidth,trim = 0mm 0mm 0mm 0mm,clip]{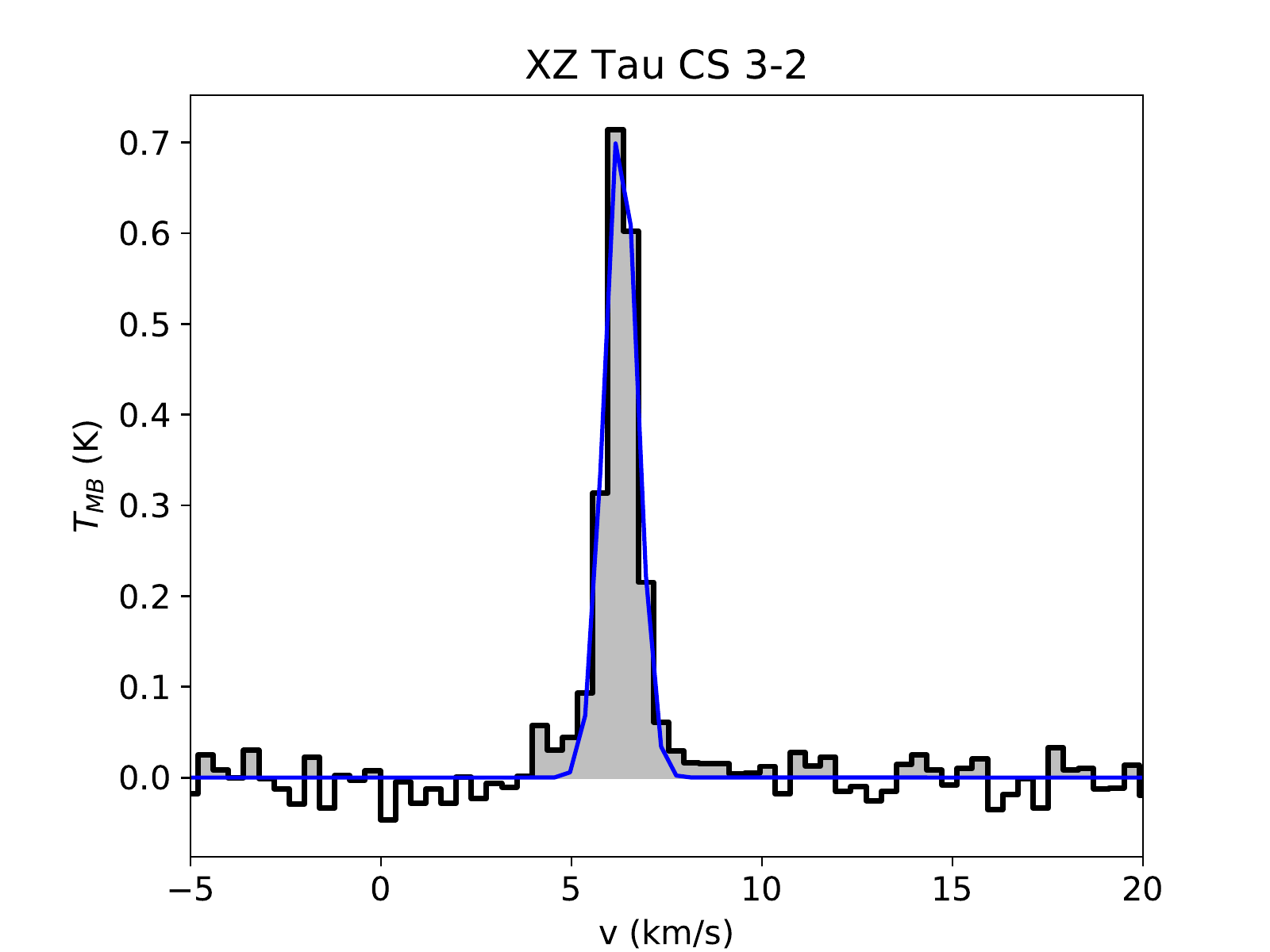}
 \includegraphics[width=0.24\textwidth,trim = 0mm 0mm 0mm 0mm,clip]{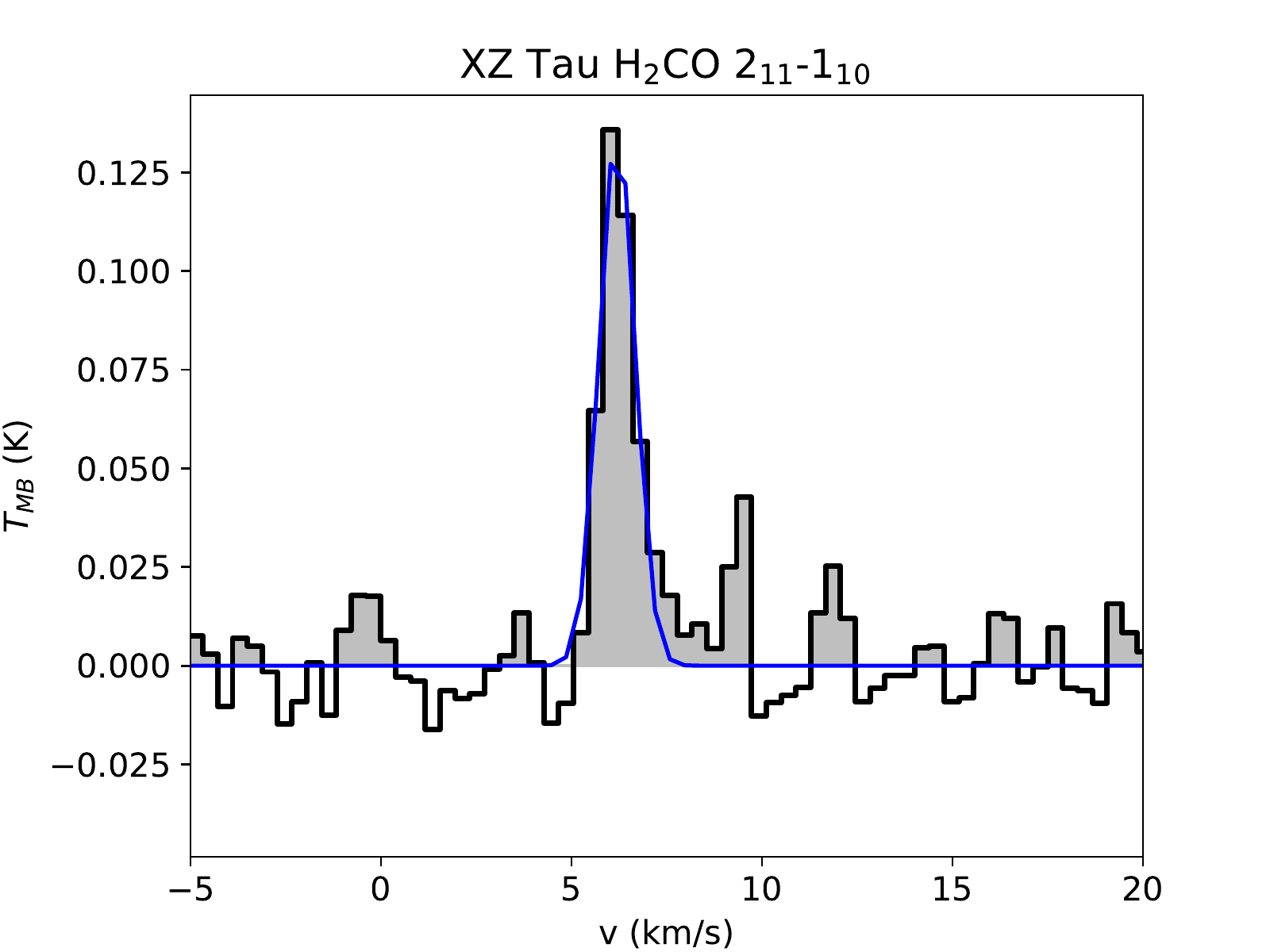}
  \caption{Spectra of sources with no H$\rm _2$S detections. The source name, molecular species, and transition are included at the top of each spectrum.}
 \label{Fig:spectra_undet_H2S}
\end{center}
\end{figure*}

Sulfur-bearing species have been detected in numerous places in the Solar System. Contrary to in the ISM, the majority of cometary detections of sulfur-bearing molecules are in the form of  H$_2$S and S$_2$ \citep{Mumma2011}. A greater diversity toward the comet Hale Bopp has been observed, including CS and SO \citep{Boissier2007}. The comets C/2012 F6  and C/2014 Q2 also contain CS \citep{Biver2016}. In situ data are available from the Rosetta mission on comet 67P/Churyumov-Gerasimenko. Using the Rosetta Orbiter Spectrometer for Ion and Neutral Analysis \citep[ROSINA;][]{Balsiger2007}, the coma has been shown to contain H$_2$S, atomic S, SO$_2$, SO, OCS, H$_2$CS, CS$_2$, and S$_2$ (and tentatively CS) gases \citep{LeRoy2015}. Furthermore, S$_3$, S$_4$, CH$_3$SH, and C$_2$H$_6$S have also been detected \citep{Calmonte2016}. Even for the large variety of S-species detected, the abundance of H$_2$S relative to H$_2$O remains around 1.5\%, similar to the limit measured in the dense ISM. 

Protoplanetary disks constitute the link between the ISM and planetary systems, and the study of sulfuretted species in these objects is of paramount importance for understanding the chemical composition of the Solar System and comets. Searches for  S-bearing molecules in protoplanetary disks have provided very few detections. So far, only one S-species, CS, has been widely detected in protoplanetary disks. The chemically related compound H$_2$CS was only detected in a disk very recently, in the transition disk MWC 480 \citep{LeGal2019a}, and was subsequently detected in two other young Class I disks, namely HL~Tau and IRAS~04302+2247 \citep{Codella2020}.  Interestingly, \cite{LeGal2019a} reported a column density ratio of CS/\ce{H2CS}$\sim$3, suggesting that the S-reservoir in disks has a larger fraction of organics than commonly thought. Recently, H$_2$S was detected in GG Tau by \cite{Phuong2018}. AB Auriga and HD 100546 remain the only transition disks with an SO detection  \citep{Pacheco2015, Booth2018}. 

In this paper we present the results of a survey of sulfuretted species toward protoplanetary disks in Taurus. In Sect. \ref{Sect:sample} we describe the sources in our sample. In Sect. \ref{Sect:observations_data_reduction} we describe how the observations were performed. In Sect. \ref{Sect:results} we present the results of our observations. A template T Tauri astrochemical model is introduced in Sect. \ref{Sect:modeling}, where we also discuss the comparison of the model with our observations. In Sect. \ref{Sect:discussion} we discuss our results in the context of sulfur chemistry in protoplanetary disks. Finally, in Sect. \ref{Sect:summary} we summarize our results.

\section{Sample \label{Sect:sample}}
The observed sample consists of nine stars in Taurus that have shown an H$\rm _2$O detection in the past. Eight of them showed  o-H$\rm _2$O 8$\rm _{18}$-7$\rm _{07}$ emission detected with the Photodetector Array Camera and Spectrometer (PACS) \citep{Riviere2012}, as well as [OI] $\rm ^{3}P_1$-$\rm ^{3}P_2$. To complete the sample we also included GV Tau, which shows o-H$\rm _2$O and p-H$\rm _2$O emission in three different transitions: 1$\rm _{10}$-1$\rm _{01}$, 1$\rm _{11}$-1$\rm _{00}$, and 2$\rm _{02}$-1$\rm _{11}$ \citep{Fuente2020}. The detections of water at far-infrared (FIR) wavelengths point to the presence of active surface chemistry, with molecular material reaching a kinetic temperature of T$_k$ $>$ 100 K. Hydrogen sulfide, similar to water, is formed on grain surfaces, and observations of its
millimeter lines would provide important information on the relevance of surface chemistry in the cold disk, thus linking the chemistry of the inner and outer disk.  The sample star positions, spectral types, and disk class are summarized in Table \ref{Tab:Sample} \citep[for a definition of classes, see][]{Lada1984, Lada1987, Andre1993}. Six out of the nine sources are Class II, one has been classified as I/II (namely, T Tau), and two, GV Tau and HL Tau, are Class I. The spectral types cover a narrow range of T Tauri types, from K0 to M2. In the following we briefly summarize the main characteristics of the sources in the sample.

\textit{AA Tau} is a K5 star \citep{Herbig1977} classified as Class II \citep{Luhman2010}. The star has a companion with a separation of 5.43$\arcsec$ \citep{Itoh2008} and is thought to harbor a powerful jet \citep{Hirth1997, Bouvier2007, Cox2013}. Its inner disk shows a rich molecular spectrum with a high abundance of simple organic molecules and water \citep{Carr2008}. The system shows $\rm [NeII]$ emission \citep{BaldovinSaavedra2012}, but the origin of the emission, whether the jet or the disk, remains unclear. However, $\rm [OI]$ and H$\rm _2$O emission at 63 $\rm \mu m$ seems to have a disk origin \citep{Riviere2012, Howard2013}. Atacama Large Millimeter Array (ALMA) observations by \cite{Loomis2017} revealed a three-ringed structure in continuum emission and the presence of an HCO$\rm ^+$ filament that connects opposite sides of the innermost parts of the disk. The authors proposed that this bridge could originate in accretion filaments that cross the disk cavity.

\begin{table}
\caption{Gaussian fits and 3$\sigma$ upper limits for sources in the sample.}
\label{Tab:line_fluxes}
\centering
\begin{tabular}{llcc}
\hline \hline
Star &Transition & Area & V$\rm _{LSR}$ \\
-- & -- & Jy km s$^{-1}$ & km s$^{-1}$ \\
\hline
GV Tau & CS 3-2 & 16.2 $\rm \pm$ 0.3 & 6.677 $\rm \pm$ 0.008 \\
             & H$\rm_2$CO 2$\rm _{11}$-1$\rm _{10}$ & 4.4 $\rm \pm$ 0.2 & 6.769 $\rm \pm$ 0.022 \\
             & H$\rm_2$S 1$\rm _{10}$-1$\rm _{01}$ & 2.3 $\rm \pm$ 0.3 & 6.702 $\rm \pm$ 0.049 \\
\hline 
HL Tau & CS 3-2 & 4.9 $\rm \pm$ 0.2 & 6.649 $\rm \pm$ 0.024\\
             & H$\rm_2$CO 2$\rm _{11}$-1$\rm _{10}$ & 1.1 $\rm \pm$ 0.1 & 6.733 $\rm \pm$ 0.108 \\
             & H$\rm_2$S 1$\rm _{10}$-1$\rm _{01}$ & 1.0 $\rm \pm$ 0.2 & 6.398 $\rm \pm$ 0.187 \\
\hline
T Tau & CS 3-2 & 11.9 $\rm \pm$ 0.1 & 8.395 $\rm \pm$ 0.009 \\
             & H$\rm_2$CO 2$\rm _{11}$-1$\rm _{10}$ &7.73$\rm \pm$ 0.09  & 8.315 $\rm \pm$ 0.012 \\
             & H$\rm_2$S 1$\rm _{10}$-1$\rm _{01}$ & 3.8 $\rm \pm$ 0.2 & 8.152 $\rm \pm$ 0.049 \\
\hline
UY Aur & CS 3-2 & 2.92 $\rm \pm$ 0.07 & 6.266 $\rm \pm$ 0.014 \\
             & H$\rm_2$CO 2$\rm _{11}$-1$\rm _{10}$ & 0.75 $\rm \pm$ 0.06 & 6.141 $\rm \pm$ 0.037\\
             & H$\rm_2$S 1$\rm _{10}$-1$\rm _{01}$ & 0.7 $\rm \pm$ 0.1 & 6.296 $\rm \pm$ 0.113 \\
\hline
AA Tau & CS 3-2 & $\rm <$0.28 & -- \\
            & H$\rm_2$CO 2$\rm _{11}$-1$\rm _{10}$ & $\rm <$0.26 & -- \\
            & H$\rm_2$S 1$\rm _{10}$-1$\rm _{01}$  & $\rm <$0.66 & -- \\
\hline
DL Tau & CS 3-2 & $\rm <$0.21 & -- \\
            & H$\rm_2$CO 2$\rm _{11}$-1$\rm _{10}$ & $\rm <$0.25 & -- \\ 
            & H$\rm_2$S 1$\rm _{10}$-1$\rm _{01}$  & $\rm <$0.78 & -- \\
\hline
FS Tau & CS 3-2 & 1.19 $\rm \pm$ 0.08 & 7.917 $\rm \pm$ 0.046 \\
            & H$\rm_2$CO 2$\rm _{11}$-1$\rm _{10}$ & 0.29 $\rm \pm$ 0.04 & 7.880 $\rm \pm$ 0.080 \\
            & H$\rm_2$S 1$\rm _{10}$-1$\rm _{01}$ & $\rm <$0.70 & -- \\
\hline
RY Tau & CS 3-2 & $\rm <$0.29 & -- \\
            & H$\rm_2$CO 2$\rm _{11}$-1$\rm _{10}$ & $\rm <$0.23 & -- \\
            & H$\rm_2$S 1$\rm _{10}$-1$\rm _{01}$ & $\rm <$0.80 & -- \\
\hline
XZ Tau & CS 3-2 & 4.0 $\rm \pm$0.1 & 6.292 $\rm \pm$ 0.012 \\
             & H$\rm_2$CO 2$\rm _{11}$-1$\rm _{10}$ & 0.79 $\rm \pm$ 0.06 & 6.197 $\rm \pm$ 0.044\\
            & H$\rm_2$S 1$\rm _{10}$-1$\rm _{01}$ & $\rm <$0.75 & -- \\
\hline
\end{tabular}
\end{table}

\textit{DL Tau} is a K7 \citep{Herbig1977} classical T Tauri star classified as Class II \citep{Luhman2010}. According to \cite{Itoh2008}, the star has a binary companion with a separation of 8.54\arcsec. It shows broad $\rm [OI]$ emission at 6300 $\rm \AA$ and $\rm [SII]$ emission at 6371 $\rm \AA$ \citep{Hartigan1995}, most likely due to the presence of an  outflow or a jet. Furthermore, He I at 10830 $\rm \AA$  is also thought to have an outflow origin \citep{Edwards2003,Edwards2006,Kwan2011}. The intensity of the $\rm [OI]$ line at 63 $\rm \mu m$ also points to the presence of an outflow \citep{Howard2013}. Continuum observations with ALMA showed a multi-ring structure, but individual rings were unresolved \citep{Long2020b}. 

\textit{FS Tau} is hierarchical triple system that consists of the close binary FS Tau A \citep[][]{Simon1992, Hartigan2003}, with a separation of 0.23-0.27$\arcsec$, and FS Tau B, with a separation of 20$\arcsec$ with respect to FS Tau A. FS Tau A members are of spectral types M3.5 and M0 \citep{Hartigan2003} and were classified as Class II \citep{Luhman2010}. According to \cite{Mundt1984}, a jet is associated with FS Tau B \citep[with M0 spectral type;][]{Luhman2010}. \cite{Riviere2016} showed that the $\rm [OI]$ profile at 63 $\rm \mu m$ is better reproduced by multiple Gaussians, indicating a contribution from different components. Water emission at 63 $\rm \mu m$ is associated with FS Tau A \citep{Riviere2012, Riviere2016}. Observations of the CO J=2-1 transition toward FS Tau A with ALMA revealed the presence of two streamers that connect the circumbinary disk with the central binary. In the remainder of the paper, when we refer to FS Tau, we are talking about FS Tau A.

\textit{GV Tau} is a binary system \citep{Leinert1989} with a separation of 1.2$\arcsec$. Below 3.8 $\rm \mu m$, GV Tau S dominates the emission, while for wavelengths larger than 4 $\rm \mu m$ the dominant contribution is the deeply obscured GV Tau N \citep{Leinert1989}. Both members of the binary have been classified as Class I. GV Tau  N was one of the first sources to be detected in HCN and C$_2$H$_2$, and the only one in CH$\rm _4$ in the near-infrared (NIR) range \citep{Gibb2007, Gibb2008, Doppmann2008, Gibb2013}. High resolution mid-infrared (MIR) spectroscopy of GV Tau N reveals a rich absorption spectrum with individual lines of C$_2$H$_2$, HCN, NH$_3$, and H$_2$O \citep{Najita2020}. \citet{Fuente2012} reported the first millimetric interferometric images of the HCN 3-2 and HCO$^+$ 3-2 lines at an angular resolution of $\sim$50 au, showing that the HCN 3$\rightarrow$2 emission only comes from GV Tau N. Based on higher-spatial-resolution millimeter images and FIR observations of $^{13}$CO, HCN, CN, and H$_2$O, \citet{Fuente2020} proposed that  GV Tau N is itself a binary in which the disk of the primary component is highly inclined relative to the circumbinary disk.

\textit{HL Tau} is a well-known K5 star \citep{White2004} classified as Class I \citep{Luhman2010} that powers an optical jet \citep{Mundt1983}. It shows strong $\rm [OI]$ emission at 63 $\rm \mu m$, most likely with a jet or outflow origin \citep{Howard2013}. However, the line profile was well reproduced by a combination of two Gaussians \citep{Riviere2016}, indicating a contribution from different components and leaving open the possibility of disk emission. ALMA observations revealed a set of rings and gaps in a highly structured protoplanetary disk \citep{ALMA2015}, which were not expected in such a young protoplanetary disk. One interpretation of such structures is that they are the result of planet formation.

\textit{RY Tau} is a K1 star \citep{Herbig1977} that hosts a Class II protoplanetary disk \citep{Luhman2010} and has a binary companion separated by 10.86$\arcsec$ \citep{Itoh2008}. The system powers a well-known jet that extends to 31$\arcsec$ from the star \citep{StOnge2008}. The jet was recently imaged in detail \citep{Garufi2019}, and the launching date of a jet spot was traced back to 2006, supporting theories of episodic accretion. Furthermore, the system shows free-free emission that is consistent with the presence of a thermal wind \citep{Rodmann2006}. ALMA observations of the continuum emission toward RY Tau \citep{Francis2020} revealed a full disk with a brightening in the innermost regions that could be due to the presence of an unresolved inner disk.

\textit{T Tau} is a hierarchical triple system. The northern component (T Tau N) is a K0 star \citep{Herbig1988}, while the southern one (T Tau S) is a deeply embedded binary  system \citep{Koresko2000} with a separation of 0.61$\arcsec$ with respect to T Tau N \citep{Dyck1982}. Both T Tau N and T Tau S are associated with jets \citep{Buehrke1986, Reipurth1997}. T Tau N has been classified as Class II, while T Tau S is a Class I system \citep{Luhman2010}. According to \cite{Lorenzetti2005}, its Infrared Space Observatory (ISO) MIR and FIR spectrum is the richest among pre-main-sequence stars. It showed strong $\rm [OI]$ and H$\rm _2$O at 63 $\rm \mu m$ \citep{Riviere2012}. Its $\rm [OI]$ line profile at 63 $\rm \mu m$ is well reproduced by a combination of a least two Gaussians, indicating a contribution from different components. However, given the complexity of the system it is hard to determine if these origins are in the jets, disks, or a combination of the two. The source was observed with ALMA \citep{Long2020}, and both the continuum from T Tau N and the the continuum from T Tau S were detected. The sources are compact, and no substructures were observed. 

\textit{UY Aur} is a binary system consisting of an M0 star and an M2.5 star \citep{Hartigan2003}, with a separation of 0.88$\arcsec$ \citep{White2001}. The system was classified as Class II \citep{Luhman2010} and is associated with a jet \citep{Hirth1997}. According to \cite{Uvarova2020}, the primary powers a wide-angle, fast wind on both sides, plus a collimated wind in the direction of the secondary, while the secondary only has a collimated jet. Observations with ALMA revealed a compact disk ($\rm \sim$ 0.4\arcsec) with no substructures \citep{Long2020}.

\textit{XZ Tau} is itself  a close binary \citep[M2+M2;][]{Hartigan2003} that belongs to a binary system, together with HL Tau, with a separation of 23.05$\arcsec$. The source has been classified as Class II \citep{Luhman2010}. The system powers a well-known optical jet \citep{Mundt1990}. 

\begin{table*}[t!]
\caption{Two-Gaussian fits to emission lines in T Tau.}
\label{Tab:TTau_multiG}
\centering
\begin{tabular}{l|ccc|ccc}
\hline \hline
                 & \multicolumn{3}{c|}{Narrow} & \multicolumn{3}{c}{Wide} \\
Transition & Area & V$\rm _{LSR}$ & FWHM  & Area & V$\rm _{LSR}$ & FWHM \\
-- & Jy km s$^{-1}$ & km s$^{-1}$ & km s$^{-1}$ & Jy km s$^{-1}$ & km s$^{-1}$ & km s$^{-1}$ \\
\hline
CS 3-2 & 3.5 $\rm \pm$ 0.3 & 8.55 $\rm \pm$ 0.01 & 0.80 $\rm \pm$ 0.04 & 9.3 $\rm \pm$ 0.3 & 8.19 $\rm \pm$ 0.03 & 2.34 $\rm \pm$ 0.07\\
H$\rm_2$CO 2$\rm _{11}$-1$\rm _{10}$ & 2.26 $\rm \pm$ 0.10 & 8.615 $\rm \pm$ 0.009 & 0.76 $\rm \pm$ 0.02 & 5.97 $\rm \pm$ 0.14 & 7.89 $\rm \pm$ 0.03 & 2.80 $\rm \pm$ 0.06 \\ 
H$\rm_2$S 1$\rm _{10}$-1$\rm _{01}$ & 0.55 $\rm \pm$ 0.43 & 8.60 $\rm \pm$ 0.09 & 0.56 $\rm \pm$ 0.53 & 3.3 $\rm \pm$ 0.4 & 7.9 $\rm \pm$ 0.2 & 2.55 $\rm \pm$ 0.25 \\
\hline
\end{tabular}
\end{table*} 

\section{Observations and data reduction \label{Sect:observations_data_reduction}}
Observations were carried out during two different runs, in October 2019 and May 2020, with the IRAM 30 m telescope. The observations were performed using the Eight MIxer Receivers (EMIR) with  Fast  Fourier  Transform  Spectrometers (FTS) 200 centered at  159.7 (lower side band, LSB) and 165.4 GHz (upper side band, USB), with a spectral resolution of 0.1953 MHz (0.34 to 0.4 km $\rm s^{-1}$ in the observed frequency range). The system temperature during the observations was in the range 269 K to 328 K.  The achieved rms noise at the relevant frequencies was: 9$\times 10^{-3}$ K to 3$\times 10^{-2}$ K at 147 GHz, with mean value $\rm \overline{RMS}$ = (1.4$\rm \pm$0.6)$\rm \times 10^{-2}$ K; 9$\times 10^{-3}$ K to 2.4$\times 10^{-2}$ K at 150 GHz, with mean value $\rm \overline{RMS}$ = (1.1$\rm \pm$0.5)$\rm \times 10^{-2}$ K; and 1.8$\times 10^{-2}$ K to 5.2$\times 10^{-2}$ K at 169 GHz, with mean value $\rm \overline{RMS}$ = (2.6$\rm \pm$1.0)$\rm \times 10^{-2}$ K. In this paper we present the results of our survey for the CS 3-2 ($\rm \nu_0 = 146.969~ GHz$), \ce{H2CO} $\rm 2_{1,1}-1_{1,0}$ ($\rm \nu_0 = 150.498~ GHz$), and \ce{H2S} $\rm 1_{1,0}-1_{0,1}$ ($\rm \nu_0 = 168.763~ GHz$) lines.  The provided numbers are given in main beam temperature ($\rm T_{MB}$) scale, computed from antenna temperature ($\rm T_{A}^{*}$) using the following values for forward and beam efficiencies: 0.93 and 0.74 ($\rm \nu_0 = 146.969~ GHz$), 0.93 and 0.73 ($\rm \nu_0 = 150.498~ GHz$ and $\rm \nu_0 = 168.763~ GHz$). The half-power beam widths (HPBWs) at these frequencies are 15.85$\arcsec$, 15.58$\arcsec$, and 14.17$\arcsec$, respectively. The data reduction was carried out using \texttt{GILDAS}\footnote{See \texttt{http://www.iram.fr/IRAMFR/GILDAS} for  more information about GILDAS software.}\texttt{/CLASS} following a standard procedure. 

The spectra baselines were fitted by applying a third degree polynomial to regions with no line emission, and they were then subtracted from the observed spectra. Line fluxes (areas) were computed by fitting 1D Gaussians to the observed spectra and using the formula of the Gaussian area. The resulting line fluxes are summarized in Table \ref{Tab:line_fluxes}. In the case of T Tau spectra, the line profiles were better reproduced using  two Gaussians, and the resulting components are summarized in Table \ref{Tab:TTau_multiG}. Three-sigma upper limits for undetected sources were derived assuming a line full width at half maximum (FWHM) of 1.4 km s$\rm ^{-1}$, the average width for detected lines.

\begin{figure*}[t!]
\begin{center}
 \includegraphics[width=0.45\textwidth,trim = 0mm 0mm 0mm 0mm,clip]{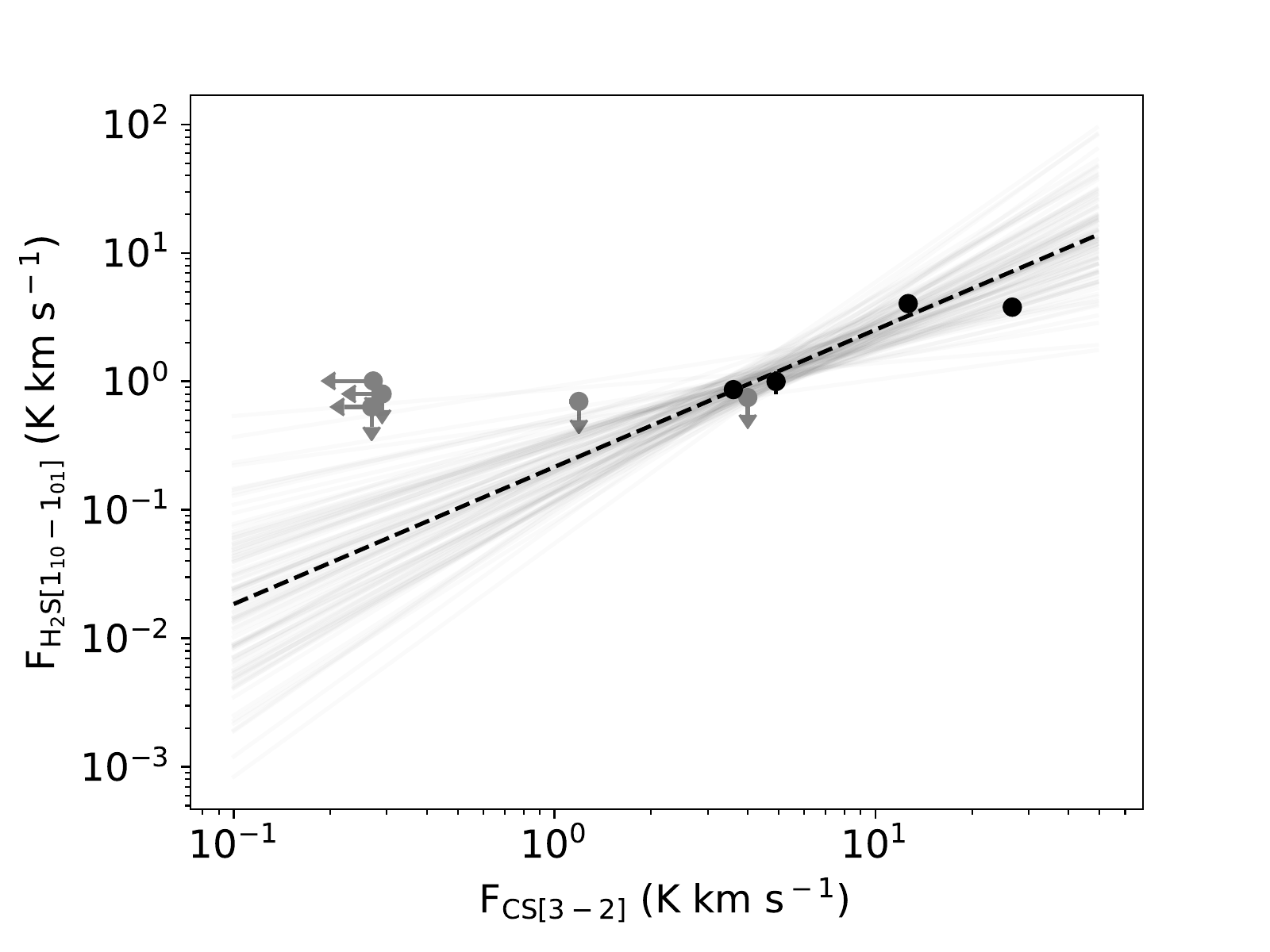}
 \includegraphics[width=0.45\textwidth,trim = 0mm 0mm 0mm 0mm,clip]{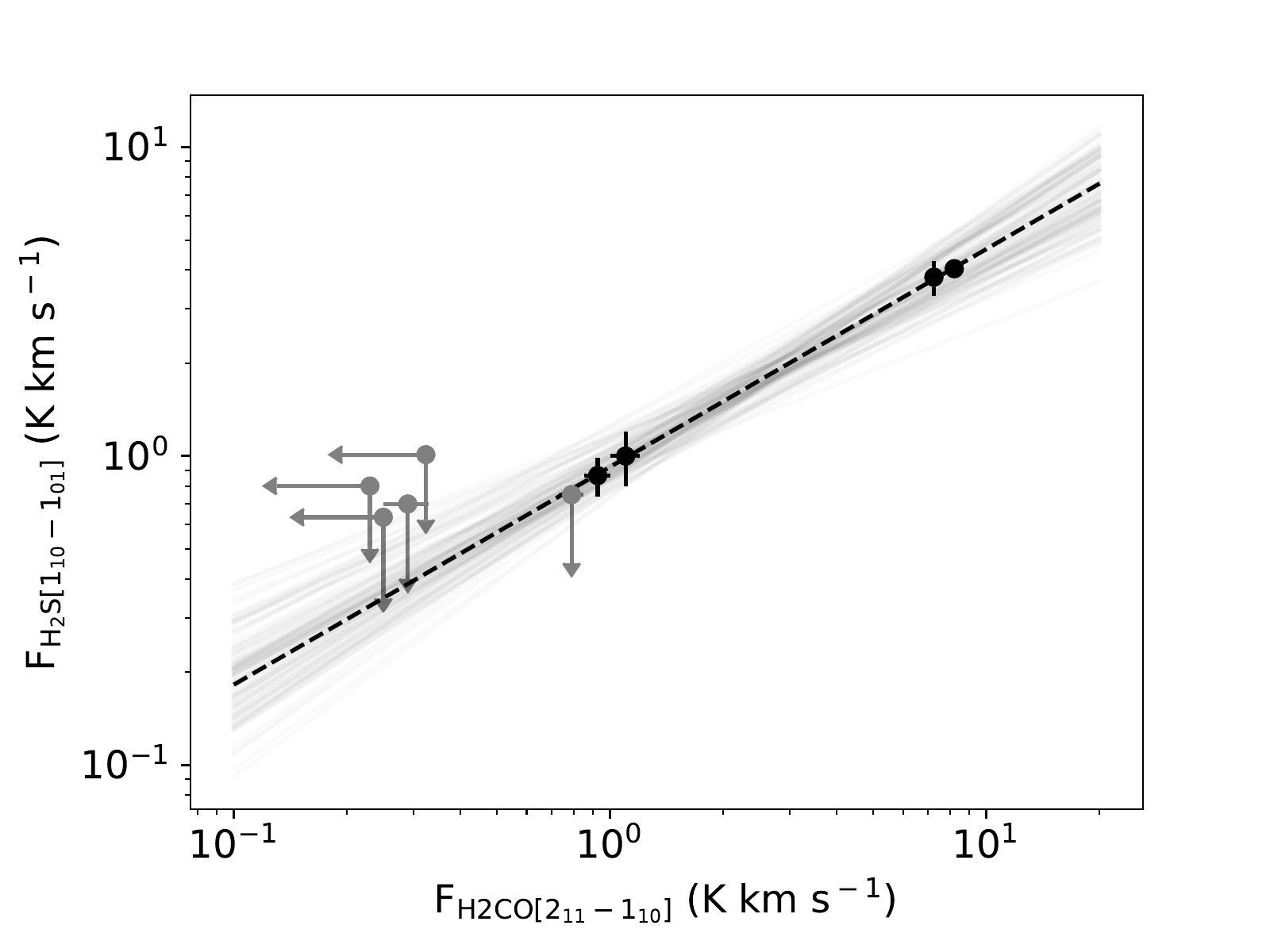}

 \includegraphics[width=0.45\textwidth,trim = 0mm 0mm 0mm 0mm,clip]{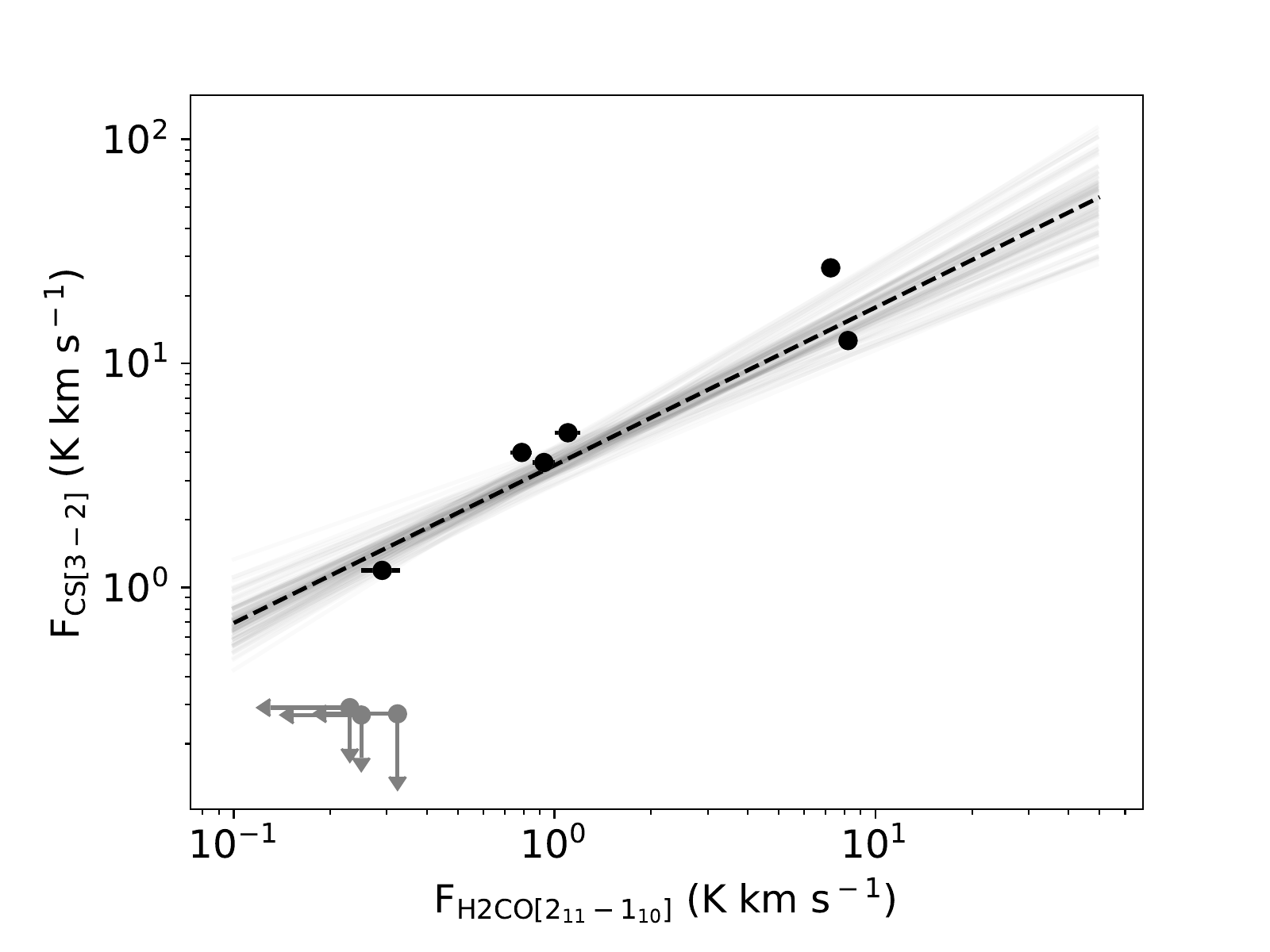} 
  \includegraphics[width=0.45\textwidth,trim = 0mm 0mm 0mm 0mm,clip]{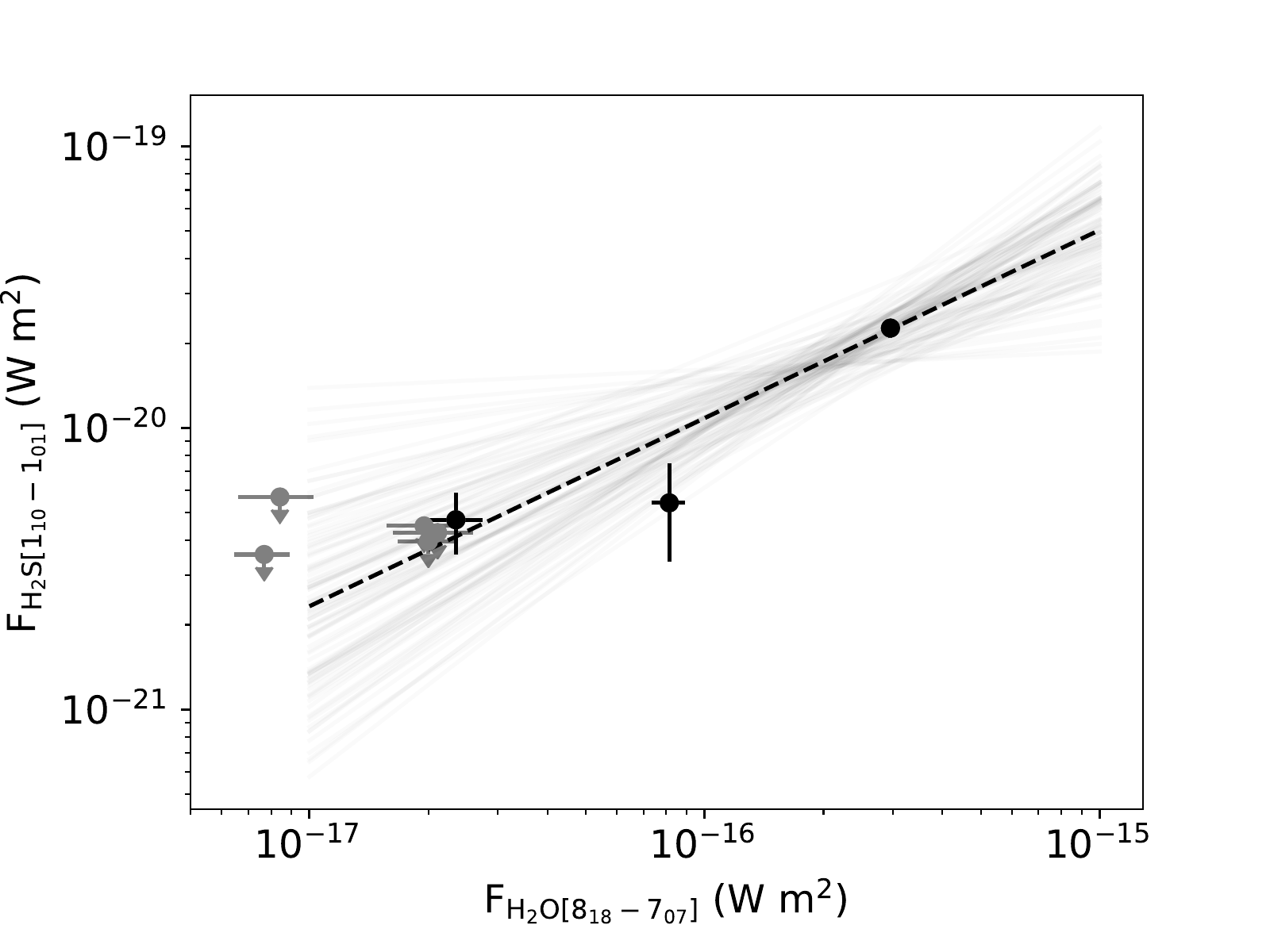} 
  \caption{Correlations between line fluxes of the different lines observed. Black dots show the position in the diagrams of detected sources, while gray arrows show the position of upper limits. The dashed black line in each plot depicts a linear fit to logarithmic-scale data. The gray lines show a random selection of 100 models drawn from the obtained posterior distributions. The line fluxes have been normalized to the distance to Taurus (140 pc) to take the dispersion in distances among Taurus members into account. }
 \label{Fig:correlations}
\end{center}
\end{figure*}

\begin{figure*}[t!]
\begin{center}
 \includegraphics[width=0.4\textwidth,trim = 0mm 0mm 0mm 0mm,clip]{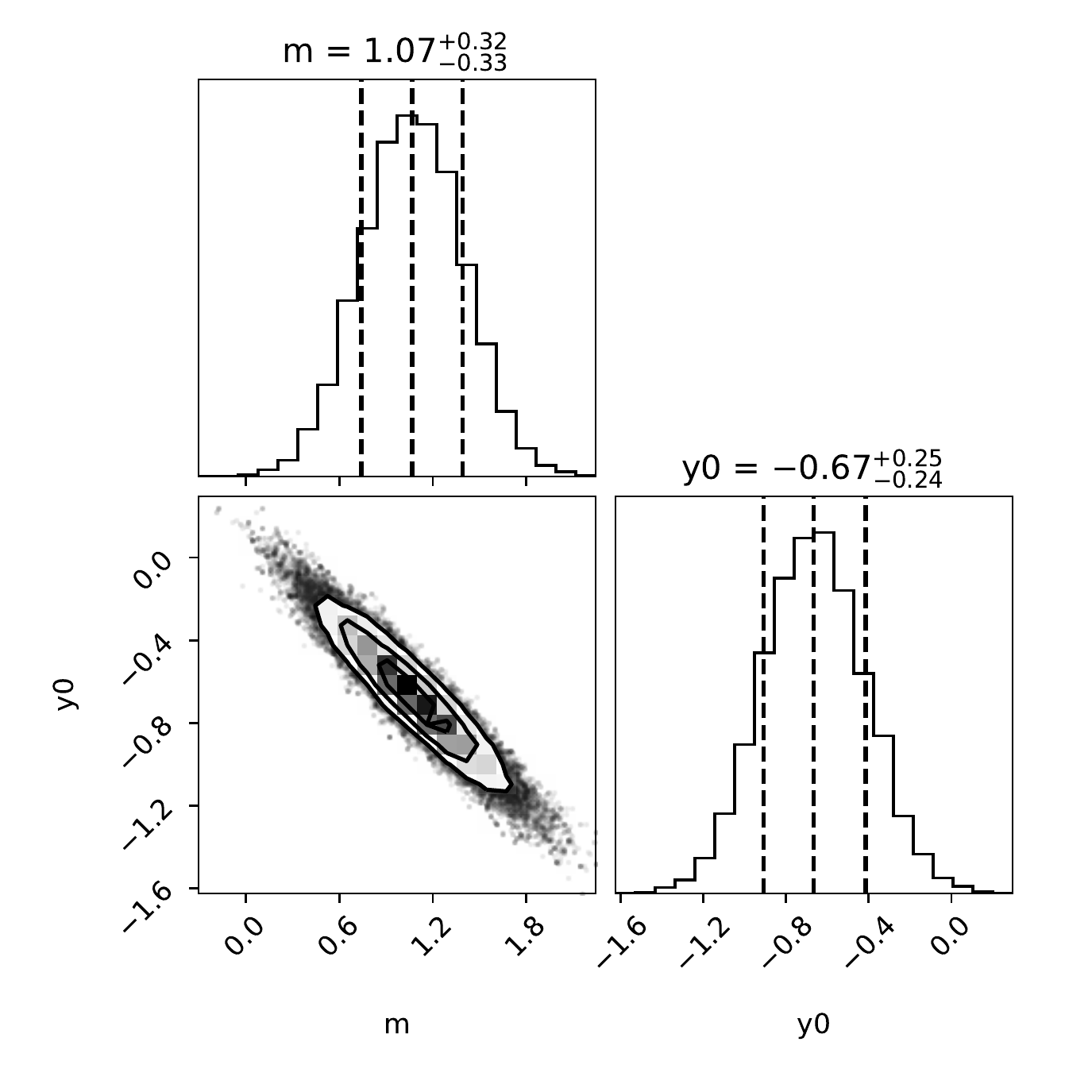}
 \includegraphics[width=0.4\textwidth,trim = 0mm 0mm 0mm 0mm,clip]{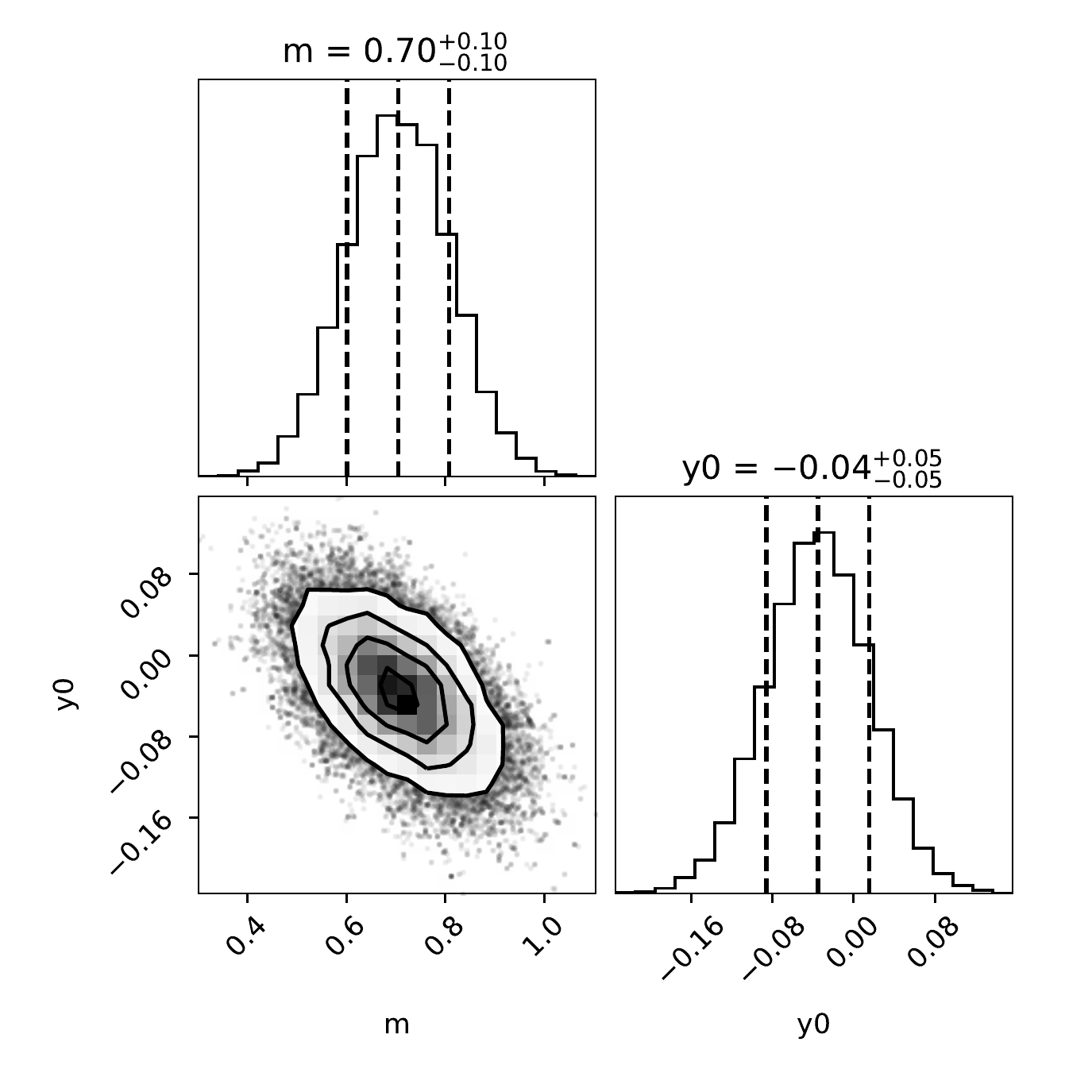}

 \includegraphics[width=0.4\textwidth,trim = 0mm 0mm 0mm 0mm,clip]{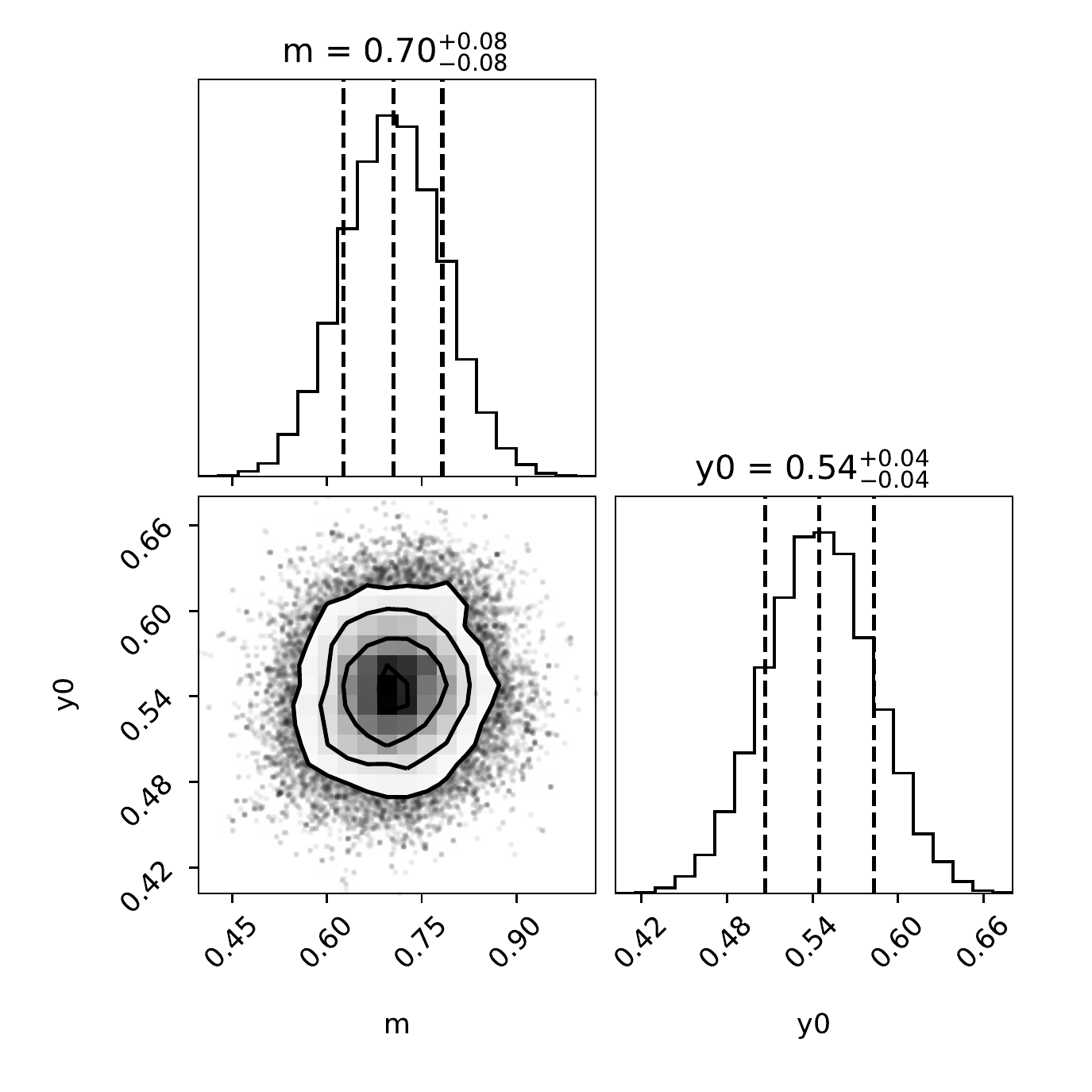} 
  \includegraphics[width=0.4\textwidth,trim = 0mm 0mm 0mm 0mm,clip]{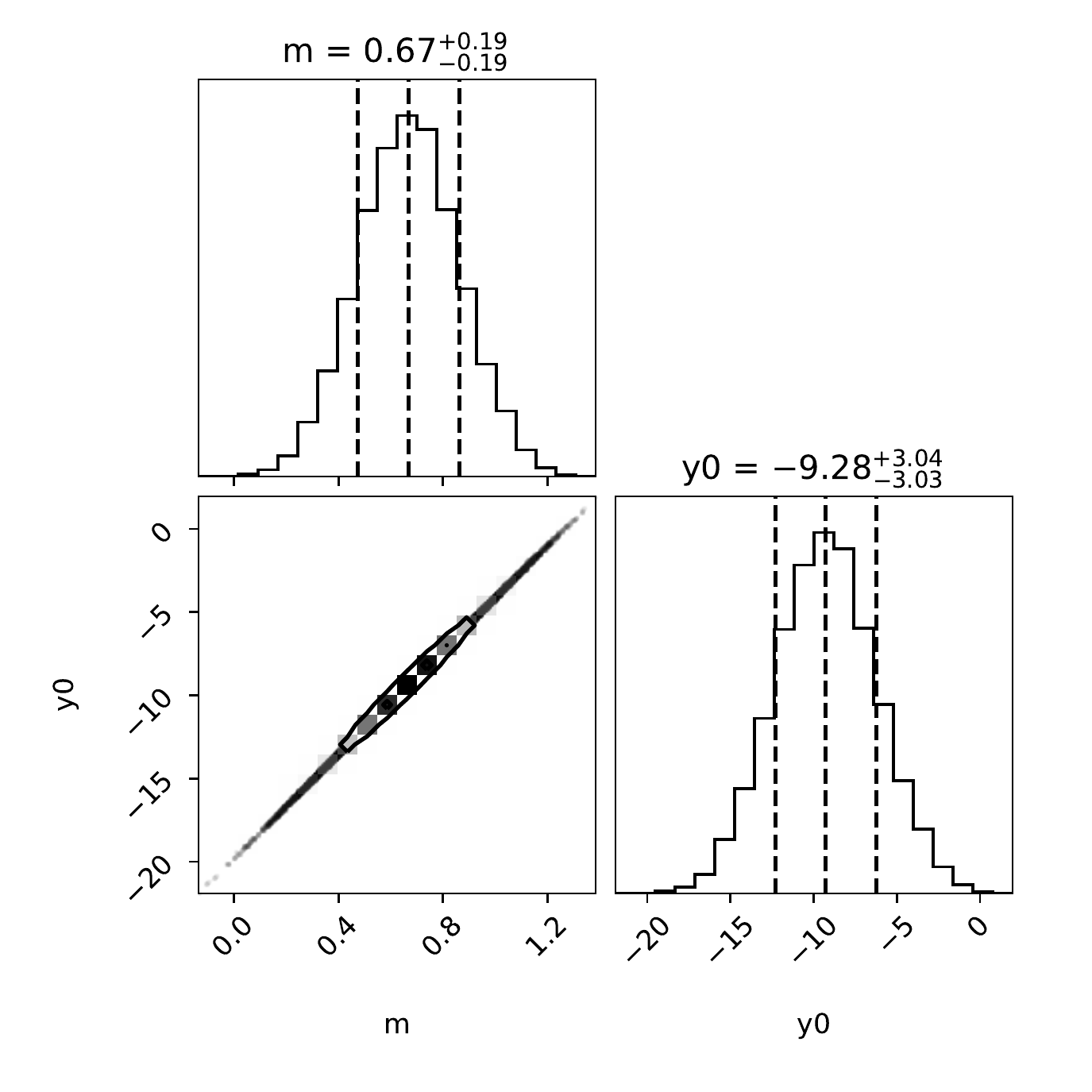} 
  \caption{ Posterior distributions for the slope (m) and the intercept ($\rm y_0$) for the different correlations explored in the logarithmic space. From left to right and top to bottom, the correlations are: \ce{H2S} vs. \ce{CS}, \ce{H2S} vs. \ce{H2CO}, \ce{CS} vs. \ce{H2CO}, and \ce{H2S} vs. \ce{H2O}. The vertical dashed lines represent the 16th, 50th, and 84th percentiles. The plots were generated using the PYTHON module CORNER \citep{Foreman2016}.}
 \label{Fig:correlations}
\end{center}
\end{figure*}

\section{Results \label{Sect:results}}
We show in Table \ref{Tab:line_fluxes} the resulting line fluxes derived from Gaussian fits for the detected transitions, as well as upper limits for undetected lines. The observed spectra are shown in Figs. \ref{Fig:spectra_det_H2S} and \ref{Fig:spectra_undet_H2S}. In all the sources the brightest line detection is CS 3-2 at 147 GHz, which was detected toward six out of nine observed sources, with line fluxes in the range 1.19 Jy km s$\rm^{-1}$ to 16.2 Jy km s$\rm^{-1}$.  The second brightest line is  o-H$\rm _2$CO 2$\rm _{11}$-1$\rm _{10}$  at 150 GHz, with line fluxes in the range 0.29 Jy km s$\rm^{-1}$ to 7.73 Jy km s$\rm^{-1}$. Formaldehyde shows the same detection ratio as CS 3-2 and was detected toward the same sources.

The most interesting result from our molecular survey is the detection of o-H$\rm _2$S 1$\rm _{10}$-1$\rm _{01}$ at 168.763 GHz in four out of the nine observed sources (GV Tau, HL Tau, T Tau, and UY Aur). The H$\rm _2$S line fluxes are in the range 0.7 Jy km s$\rm^{-1}$ to 3.8 Jy km s$\rm^{-1}$. Previously, H$\rm _2$S emission had only been detected toward one Class II object, GG Tau \citep{Phuong2018}, despite being searched for in another four systems, GO Tau, MWC 480, DM Tau, and LkCa 15 \citep{Dutrey2011}; in these searches, the authors used DiskFit \citep{Pietu2007} to estimate upper limits in H$\rm _2$S column densities on the order of a few 10$\rm ^{11}~cm^{-2}$.

\begin{table*}[h!]
\caption{Column densities from RADEX assuming a temperature of 20~K and a gas density of 10$\rm ^{7}~cm^{-3}$.}
\label{Tab:CD}
\centering
\begin{tabular}{lcccccc}
\hline \hline
Star & N(o-H$\rm_2$S) & N(o-H$\rm_2$S)* & N(CS) & N(CS)* & N(\ce{o-H2CO})  & N(\ce{o-H2CO})*   \\
-- & cm$\rm ^{-2}$ & cm$\rm ^{-2}$ & cm$\rm ^{-2}$ & cm$\rm ^{-2}$ & cm$\rm ^{-2}$ & cm$\rm ^{-2}$  \\
\hline
AA Tau & $\rm <$6.3$\rm \times 10^{11}$  & $\rm <$ 2.6$\rm \times 10^{12}$  &  $\rm <$1.9$\rm \times 10^{11}$  &  $\rm <$9.7$\rm \times 10^{11}$  & $\rm <$2.1$\rm \times 10^{11}$  &  $\rm <$9.8$\rm \times 10^{11}$ \\
DL Tau & $\rm <$7.4$\rm \times 10^{11}$ & $\rm <$3.0$\rm \times 10^{12}$  &   $\rm <$1.4$\rm \times 10^{11}$  &   $\rm <$7.0$\rm \times 10^{11}$  &  $\rm <$2.0$\rm \times 10^{11}$  &  $\rm <$9.5$\rm \times 10^{11}$ \\
FS Tau & $\rm <$6.7$\rm \times 10^{11}$ & $\rm <$2.7$\rm \times 10^{12}$  &  $\rm (6.4 \pm 0.6) \times 10^{11}$  &  $\rm (3.3 \pm 0.4) \times 10^{12}$  &  $\rm (2.2 \pm 0.4)\times 10^{11}$  &  $\rm (1.1 \pm 0.2) \times 10^{12}$ \\
GV Tau & $\rm (2.2 \pm 0.3)\times 10^{12}$  &  $\rm (9.2 \pm 2.0) \times 10^{12}$  &  $\rm (9.4 \pm 0.2) \times 10^{12}$  &  $\rm (7.4 \pm 0.2) \times 10^{13}$  &  $\rm (3.5 \pm 0.2)\times 10^{12}$  &  $\rm (1.8 \pm 0.1) \times 10^{13}$ \\
HL Tau & $\rm (9.0 \pm 4.0) \times 10^{11}$  &  $\rm (3.7 \pm 1.5) \times 10^{12}$  &  $\rm (2.7 \pm 0.2) \times 10^{12}$  &  $\rm (1.4 \pm 0.1) \times 10^{13}$  &  $\rm (8.4 \pm 0.2) \times 10^{11}$  &  $\rm (4.0 \pm 0.8) \times 10^{12}$ \\
RY Tau & $\rm <$7.6$\rm \times 10^{11}$ & $\rm <$3.1$\rm \times 10^{12}$ & $\rm <$2.0$\rm \times 10^{11}$ & $\rm <$9.7$\rm \times 10^{11}$  &  $\rm <$1.9$\rm \times 10^{11}$  &  $\rm <$8.7$\rm \times 10^{11}$ \\
T Tau & $\rm (3.7 \pm 0.4) \times 10^{12}$  &  $\rm (1.5 \pm 0.2) \times 10^{13}$  &  $\rm (6.6 \pm 0.1) \times 10^{12}$  &  $\rm (3.9 \pm 0.1) \times 10^{13}$  &  $\rm (6.0 \pm 0.1) \times 10^{12}$  &  $\rm (3.1 \pm 0.1) \times 10^{13}$ \\
UY Aur & $\rm (6.5 \pm 2.0) \times 10^{11}$  &  $\rm (2.6 \pm 0.7) \times 10^{12}$  &  $\rm (1.6 \pm 1.0) \times 10^{12}$  &  $\rm (8.4 \pm 0.4) \times 10^{12}$  &  $\rm (6.0 \pm 0.8) \times 10^{11}$  &  $\rm (2.8 \pm 0.3) \times 10^{12}$ \\
XZ Tau & $\rm <$7.2$\rm \times 10^{11}$  &  $\rm <$2.9$\rm \times 10^{12}$  &  $\rm (2.2 \pm 0.1) \times 10^{12}$  &  $\rm (1.2 \pm 0.1) \times 10^{13}$  &  $\rm (6.1 \pm 0.6) \times 10^{11}$  &  $\rm (2.9 \pm 0.3) \times 10^{12}$ \\

\hline
\end{tabular}
\tablefoot{$\rm ^*$ Computed after correcting for beam dilution, applying beam dilution factors of 0.25 (\ce{H2S}), 0.20 (CS), and 0.21 (\ce{H2CO}) to main beam temperatures.}
\end{table*} 

As mentioned in Sect. \ref{Sect:observations_data_reduction}, the lines detected toward T Tau can be better reproduced using a combination of two Gaussians, rather than only one. The components are a narrow one, with $\rm \overline{FWHM}$ = (0.7 $\rm \pm$ 0.1) km s$\rm ^{-1}$, and a wide one, with $\rm \overline{FWHM}$ = (2.6 $\rm \pm$ 0.2) km s$\rm ^{-1}$. The parameters of the resulting fits are summarized in Table \ref{Tab:TTau_multiG}. The area below the wide component is always larger than than the area of the narrow component. For CS and o-H$\rm _2$CO, the ratios are 2.7$\rm \pm$0.3 and 2.6$\rm \pm$0.2, respectively, while for o-H$\rm _2$S the ratio is 6.0$\rm \pm$ 5.4. Although the ratio is larger for H$\rm_2$S, the low S/N\ of the narrow component precludes any further conclusions. Given the complexity of the system, which is a hierarchical triple system (see Sect. \ref{Sect:sample}), we cannot assert whether the components are due to different physical phenomena (winds, disk, envelope) or are due to emission from different members of the hierarchical triple system. High-spatial-resolution images are needed to tackle this question. We also observe  high-velocity wings in the CS spectra of GV Tau, and tentatively in UY Aur. However, models using only one Gaussian provide a good fit to the overall profile.

The observed line profiles provide some hints about the origin of the emission of these molecules. All the targets considered are associated with jets. It is well known that the abundances of H$_2$CO and S-bearing molecules can be enhanced in the shocks produced when the jets impact the molecular cloud. Indeed, these molecules present intense emission in the high-velocity wings of the bipolar outflows associated with Class 0 sources such as L~1157 \citep{Bachiller1997, Holdship2019}. However, the spectra observed in our sample of Class I and II objects present line widths of $\rm < 3~km~s^{-1}$, which is more consistent with the emission arising in the circumstellar disk.  We cannot rule out that a fraction of the emission comes from the remnant envelope in  the Class I protostars GV Tau and HL Tau or from the envelope around T Tau S.  Also, there may be some contribution from the envelope associated with T Tau S. However, this is not expected in the case of UY Aur, which is a Class II object.

We show in Fig. \ref{Fig:correlations} the correlations between the different observed line fluxes. In  Fig. \ref{Fig:correlations} we also include the relation between o-\ce{H2S} and \ce{H2O} at 63 $\rm \mu m$ from \citet{Riviere2012}. The discussion of these correlations is hampered by the small size of the sample, but we identify  interesting trends:  o-\ce{H2CO} and  o-\ce{H2S} seem to be correlated, and there seems to be a strong correlation between  o-\ce{H2S} and o-\ce{H2O}. There are also hints of a correlation between  o-\ce{H2S} and CS as well as between CS and o-\ce{H2CO}. We emphasize that the correlations discussed in this paragraph are drawn from a small sample and that a larger sample is needed before they can be firmly established. To illustrate the uncertainty in the observed correlations, we repeated the linear fit process ($\rm y=y_0 + m x$) using a Bayesian approach, by means of the affine invariant Markov chain Monte Carlo \citep[MCMC;][]{Goodman2010} implemented in the Python package \textit{emcee} \citep{Foreman2013}, using 50 walkers, 2000 steps, and a burning parameter of 1000 (meaning that the first 1000 steps are removed from the chain to perform the statistics). The fits were performed in logarithmic space. A random selection of 100 fits for each correlation is included in Fig. \ref{Fig:correlations} as gray lines. We used the medians of the distribution as  proxies for the best fit parameters, and we used the 16th and 84th percentiles as a measure of the uncertainties in these parameters. The slope and intercept of the \ce{H2S} versus \ce{H2CO} diagram are m = 0.7$\rm \pm$0.1 and $\rm y_0 = -0.04\pm 0.05$ (i.e., relative uncertainties of 14\% and 125 \%); for \ce{H2S} versus \ce{CS} we get m = 1.07$\rm ^{+0.32}_{-0.33}$ and $\rm y_0 = -0.67 ^{+0.25}_{-0.24}$ (relative uncertainties of 30\% and 37 \%); for \ce{CS} versus \ce{H2CO} we get m = 0.7$\rm \pm 0.08$ and $\rm y_0 = 0.54 \pm 0.04$ (relative uncertainties of 11\% and 7\%); and for \ce{H2S} versus \ce{H2CO} we get m = 0.67$\rm \pm 0.19$ and $\rm y_0 = -9.3 \pm 3.0$ (relative uncertainties of 28\% and 33\%).

\subsection{Column densities \label{Sect:col_dens}}
We used RADEX \citep{vanDerTak2007} to derive molecular column densities assuming a characteristic disk kinetic temperature of 20~K \citep{Williams2011} and a gas density of 10$\rm ^{7}~cm^{-3}$. In our first trial, we assumed the beam filling factor to be equal to 1 (i.e., the emission is filling the beam). However, these column densities are lower limits to actual values since no beam dilution was applied. Assuming that most of the emission comes from the disk and assuming a typical disk radius of r = 500~au in T Tauri stars \citep{Pegues2020,Phuong2018,LeGal2019b}, at the distance to Taurus \citep[140 pc;][]{Kenyon2008} the resulting beam dilution factors are 0.25 for o-\ce{H2S}, 0.21 for  o-\ce{H2CO}, and 0.20 for CS. If such factors are applied, we derive \ce{o-H2S} column densities in the range 2.6$\rm \times$10$\rm ^{12}~cm^{-2}$ to 1.5$\rm \times 10^{13}~cm^{-2}$, which are comparable to the column density derived in GG Tau  by \cite{Phuong2018} and one to two orders of magnitude larger than the upper limits derived by \cite{Dutrey2011}. We note that our estimates of the column densities are based on single-dish observations, while the value derived by \cite{Phuong2018} is based on interferometric observations. As such, the comparison relies on our source size assumption and is subject to large uncertainties. Upper limits derived for \ce{o-H2S} column densities are in the range 2.6$\rm \times 10^{12}~cm^{-2}$ to 3.1$\rm \times 10^{12}~cm^{-2}$, approximately one order of magnitude larger than the upper limits derived by \cite{Dutrey2011}.

\begin{figure*}[t!]
\begin{center}
 \includegraphics[width=0.9\textwidth,trim = 0mm 0mm 0mm 0mm,clip]{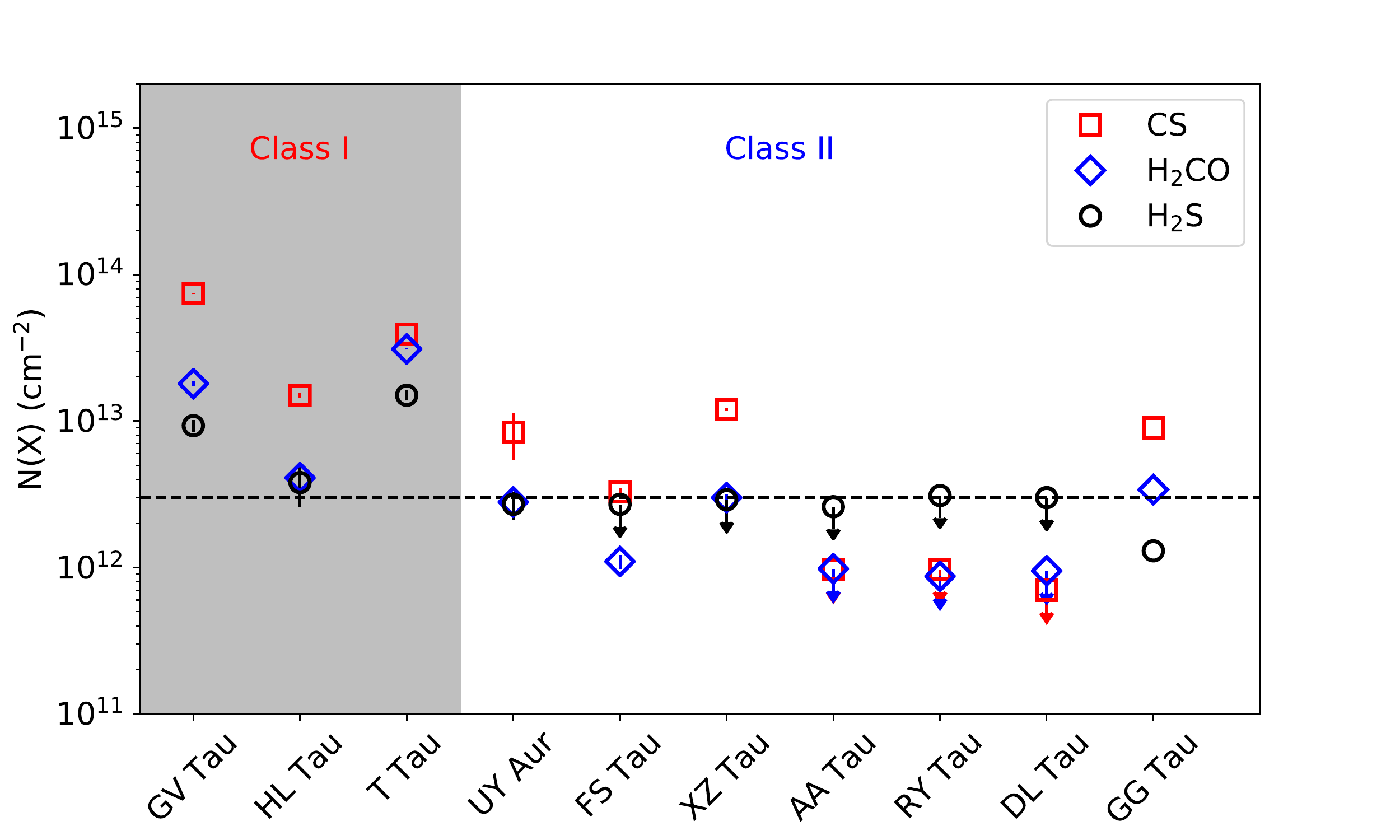}
  \caption{\ce{CS}, \ce{H2CO}, and \ce{H2S} column densities for sources in the sample. We show column densities for GG Tau as well. }
 \label{Fig:survey_col_dens}
\end{center}
\end{figure*}

The resulting column densities are shown in Table \ref{Tab:CD}, and in Fig. \ref{Fig:survey_col_dens} we show the column densities of the different species for the sources in the sample computed assuming beam dilution. The Class I sources, GV Tau and HL Tau, show the highest column densities. T Tau shows intermediate values. UY Aur, FS Tau, AA TAu, RY Tau, and DL Tau are Class II sources and show smaller column densities. XZ Tau is a Class II, but the computed column densities are larger than the other Class II sources in the sample. Contamination from HL Tau may be affecting the measured line fluxes and, subsequently, the column densities. Indeed, as we show in the following section, the computed column densities toward XZ Tau are very close to the values derived for HL Tau. The separation between HL Tau and XZ Tau is 23.05$\arcsec$, and the beam HPBWs at 168.76 GHz, 150.5 GHz, and 146.97 GHz are 14.2$\arcsec$, 15.6$\arcsec$, and 15.8$\arcsec$, respectively. The contribution from HL Tau to the flux by XZ Tau would be between 0.1$\%$ at 168.76 GHz  and 0.3$\%$ at 146.97 GHz. Therefore, it is unlikely that the fluxes measured toward XZ Tau are due to contamination by HL Tau. High-spatial-resolution observations are needed to resolve this issue.
 
The derived values depend on the values assumed for the temperature and density. We assumed that the disk is in local thermodynamical equilibrium (LTE) and fixed the density to a value large enough to guarantee LTE conditions ($\rm n_H = 10^7~cm^{-3}$). To test the impact of changes in the temperature on the derived column densities, we computed the differences with the column densities derived assuming T = 10 K and T = 30 K. The impact of such a change in temperature is small. CS column densities show an average change of 8\% when T = 30 K, and of 25\% when  T = 10 K. \ce{The o-H2CO} column densities show an average change of 21\% when T = 30 K, and of 8\% when  T = 10 K.  Finally, \ce{o-H2S} column densities change by 4\% on average when T = 30 K, and 27\% when T = 10 K.

In this section we have assumed that all the molecular emission is coming from the circumstellar disk. This is fully justified for Class II sources that have already dispersed their envelopes. However, this assumption could be more uncertain in the case of Class I sources which still retain an optically thin envelope. In order to gain insight into this problem, we compared interferometric observations of CS and H$_2$CO millimetric lines in Class 0 and I sources \citep{Sakai2014,Oya2014,Garufi2021}. While the CS and H$_2$CO emissions present important contributions from the dense envelope in the prototypical Class 0 object L1527 \citep{Sakai2014}, the emission of millimeter lines of CS and H$_2$CO with similar excitations conditions to those presented in this paper mainly comes from the circumstellar disks in Class I protostars \citep{Garufi2020, Garufi2021}. Therefore, we consider our assumption to be reasonable for our sample.

\begin{table*}[t!] \label{Table:grid_setup}
\caption{ Parameters of models in the grid.}
\begin{center}
\begin{tabular}{ll|c}
\hline\hline
Parameter  & Units & Values \\
\hline
$\rm T_{gas}$ & $\rm [K]$ & 10, 15,  20, 25, 30, 35, 40, 45 \\
$\rm n_{gas}$ & $\rm [cm^{-3}]$ & 1$\rm \times 10^{4}$, 3$\rm \times 10^{4}$, 1$\rm \times 10^{5}$, 3$\rm \times 10^{5}$, 
                                                    1$\rm \times 10^{6}$, 3$\rm \times 10^{6}$, 1$\rm \times 10^{7}$, 3$\rm \times 10^{7}$,
                                                    1$\rm \times 10^{8}$, 3$\rm \times 10^{8}$ \\
$\rm A_v $ & $\rm [mag]$ & 10 \\
$\rm f_{UV} $ & $\rm [Draine~ units]$ & 1 \\
\hline
\end{tabular}
\end{center}
\end{table*}%

\begin{table}\label{Table:initial_abundances}
\centering
\caption{Initial elemental abundances for models in the grid.}
% \begin{tabular}{@{\hskip3pt}c@{\hskip3pt} l c c @{\hskip1pt}c@{\hskip1pt}}
\begin{tabular}{lcc}
\hline\hline
Species & $n_i/n_{\text{H}}$   \\
\hline
H$_2$   & 0.5\\
He      & 9.0$\times$10$^{-2}$              \\
N       & 6.2$\times$10$^{-5}$              \\  
O       & 2.4$\times$10$^{-4}$             \\ 
C$^+$   & 1.7$\times$10$^{-4}$, 2.4$\times$10$^{-4}$  \\
S$^+$   & 1.5$\times$10$^{-5}$, 8.0$\times$10$^{-8}$\\
P$^+$   & 2.0$\times$10$^{-10}$           \\
Cl$^+$  & 1.0$\times$10$^{-9}$ \\
F$^+$   & 6.7$\times$10$^{-9}$  \\
Si$^+$  & 8.0$\times$10$^{-9}$  \\
Fe$^+$  & 3.0$\times$10$^{-9}$  \\
Na$^+$  & 2.0$\times$10$^{-9}$  \\
Mg$^+$  & 7.0$\times$10$^{-9}$  \\
\hline
\end{tabular}

\end{table}

\section{Astrochemical modeling}
\label{Sect:modeling}
 Aiming to understand the observed correlations, we produced a grid of 0D \texttt{Nautilus (v.1.1)}  models \citep{Ruaud2016,Wakelam2017} that can be compared to our observations. In the following, we summarize the properties of the models and compare the column densities computed in Sect. \ref{Sect:col_dens} with the model grid output. Our goal is not to produce a model that fits our data, but rather to synthesize a population that reproduce the observed trends.

\subsection{Model grid description}
The \texttt{Nautilus} astrochemical code \citep{Semenov2010, Loison2014, Wakelam2014, Reboussin2015} computes the evolution of molecular abundances for a given set of elemental initial abundances and physical parameters, including gas and dust temperature, gas density, cosmic ray ionization rate, and extinction. The model includes gas-phase, gas-grain, and surface chemistry reactions. \texttt{Nautilus (v.1.1)} \citep{Ruaud2016,Wakelam2017} is a refinement of previous versions that includes both grain surface and grain mantle reactions. A more detailed description of the code is available in \citet{Ruaud2016} and \citet{Wakelam2017}.

To compute our models we included both photodesorption and chemical desorption. Our first grid of models consisted of 1600 models that covered a combination of gas densities, gas temperatures, UV flux, and visual extinction. Since the goal of \texttt{Nautilus (v.1.1)} is to follow chemical evolution, it allows for the computation of the molecular abundances at different ages. However, we decided to fix the age to 1 Myr, the value typically assumed for protoplanetary disks in Taurus. A parameter that is known to impact the chemistry of protoplanetary disks is C/O. We computed two different grids using the solar C/O = 0.7 and the carbon rich C/O = 1.0. Since the species of interest are sulfur-bearing, we produced two sets of models: one with solar sulfur abundance ($\rm [S/H]~=~1.5\times 10^{-5}$) and one with depleted sulfur abundance ($\rm [S/H]~=~8\times 10^{-8}$). Varying the UV flux and visual extinction only results in larger scatter, without improving our understanding of the underlying correlations, so we decided to fix both parameters and vary the gas density and temperature. We assumed that the dust temperature is the same as the gas temperature. The cosmic ray ionization rate is $\rm \zeta=10^{-17}~s^{-1}$.  A summary of the physical parameters covered by the grid can be found in Table \ref{Table:grid_setup}. The initial elemental abundances assumed to compute the grid are summarized in Table \ref{Table:initial_abundances} .

\begin{figure*}[t!]
\begin{center}
 \includegraphics[width=0.48\textwidth,trim = 0mm 0mm 0mm 0mm,clip]{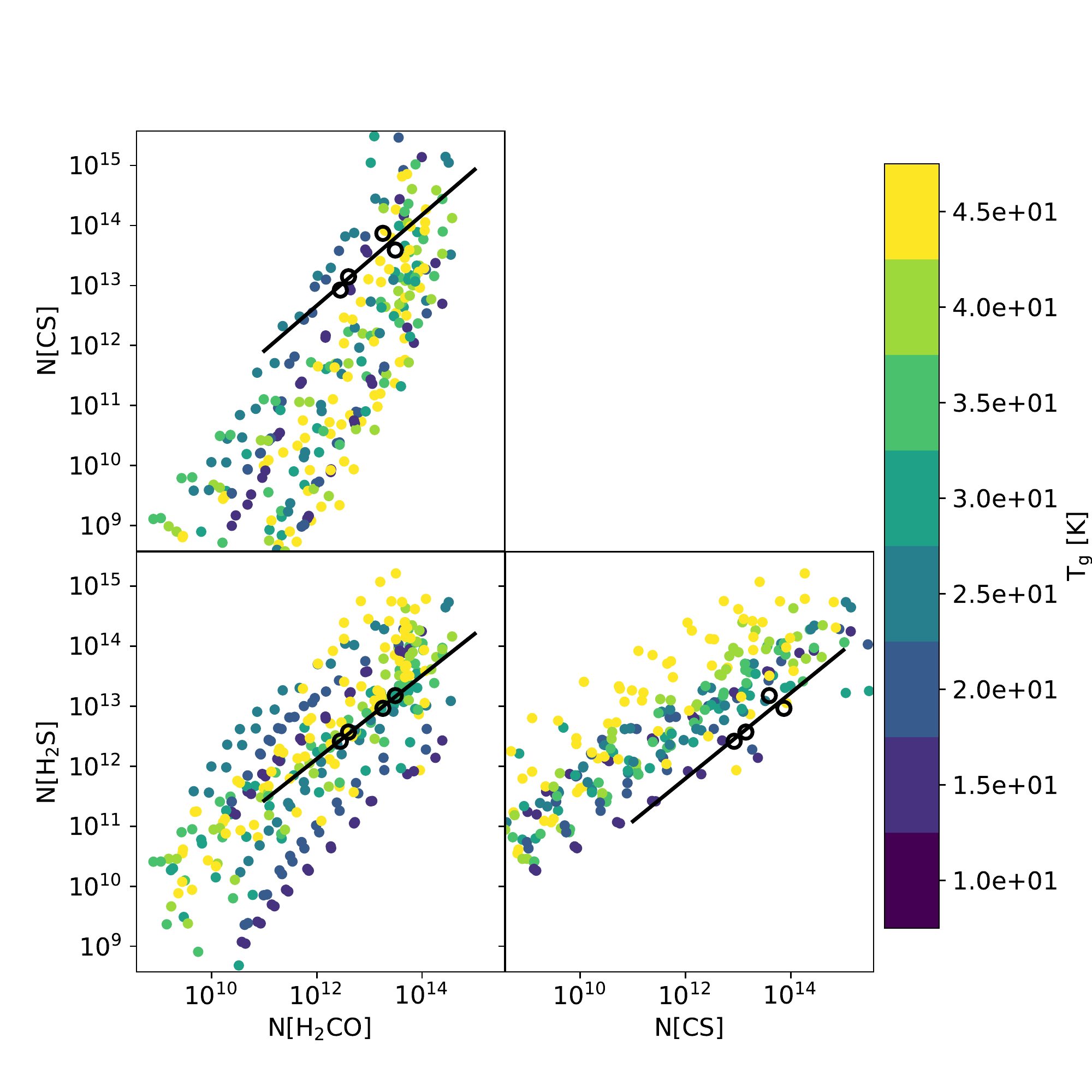} \includegraphics[width=0.48\textwidth,trim = 0mm 0mm 0mm 0mm,clip]{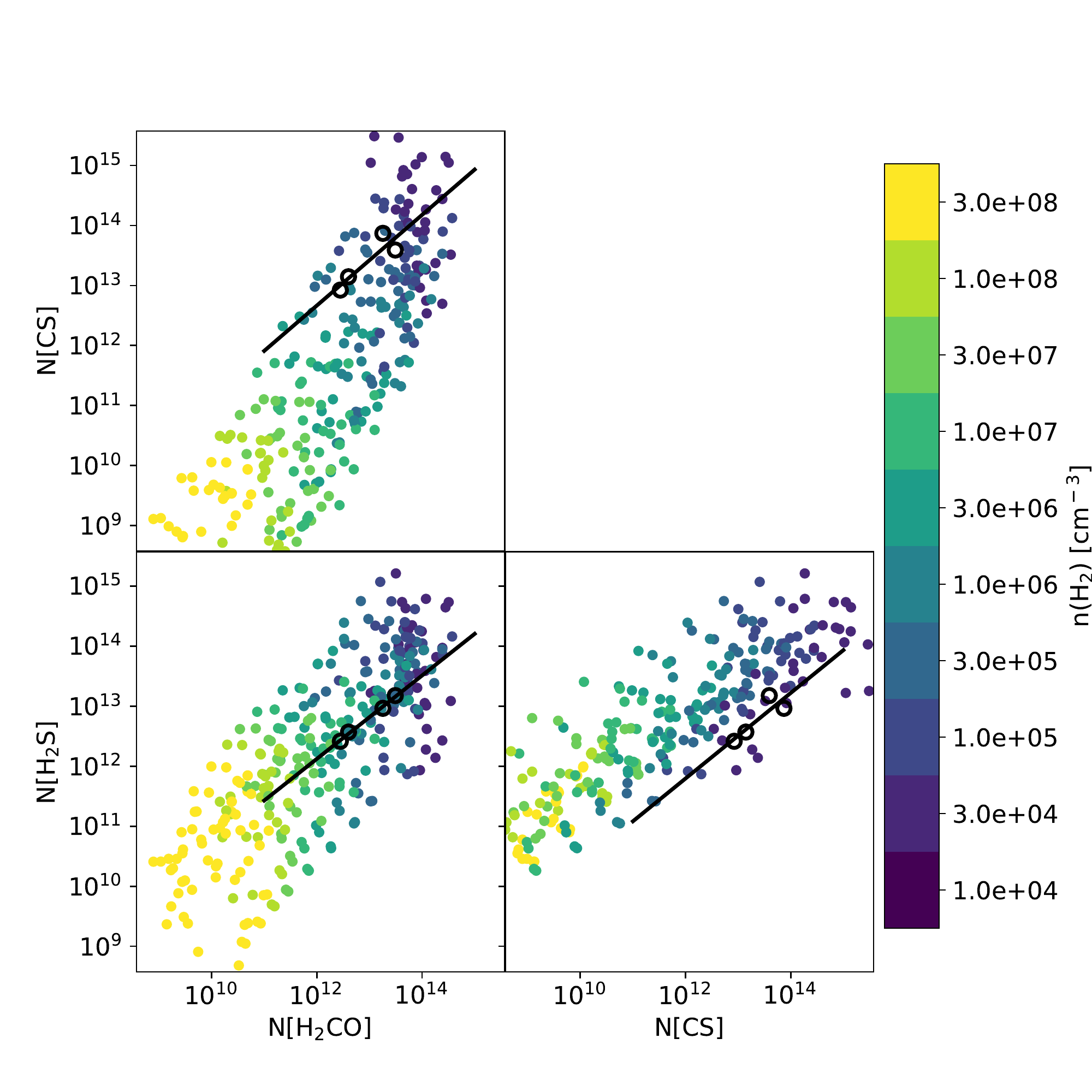}

  \includegraphics[width=0.48\textwidth,trim = 0mm 0mm 0mm 0mm,clip]{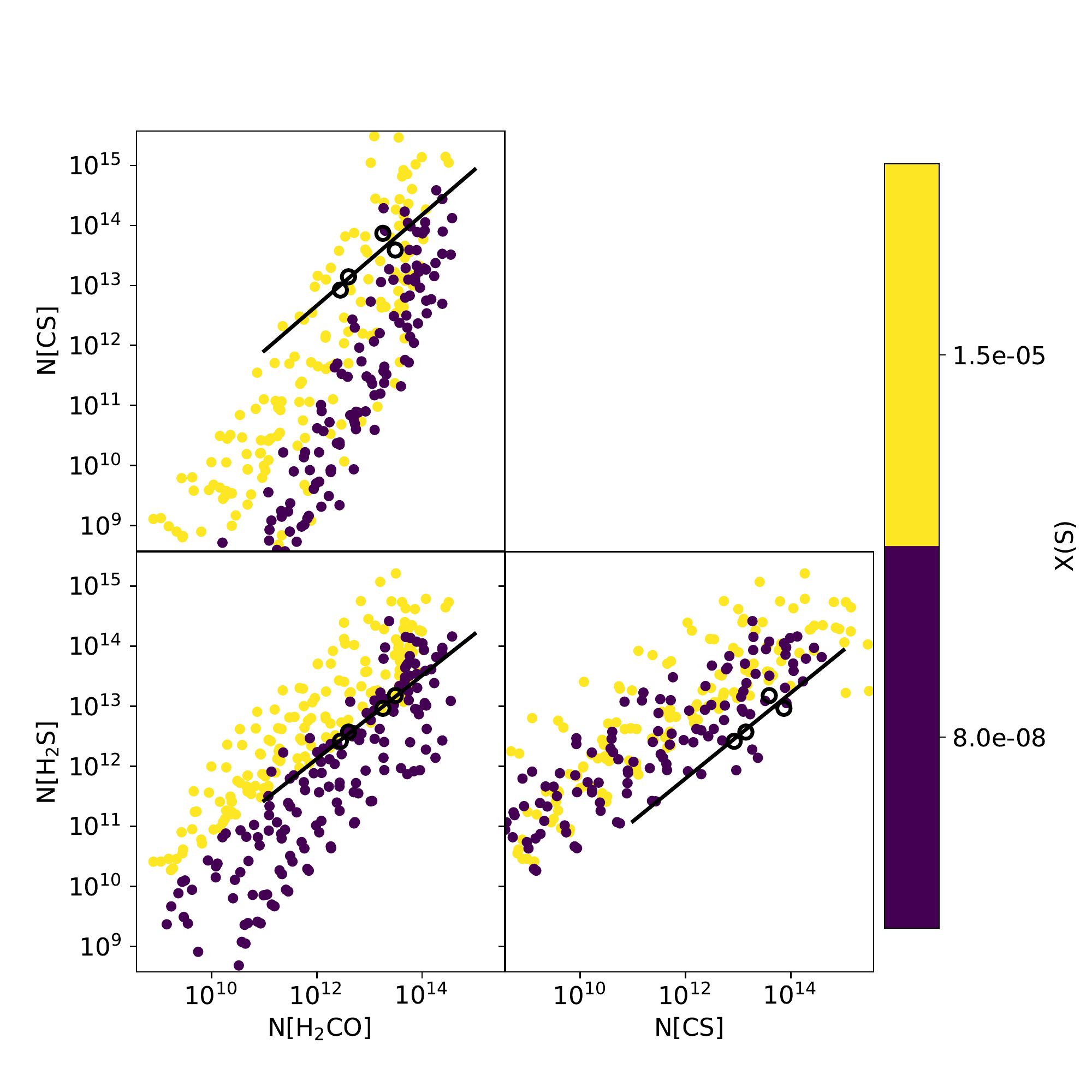} \includegraphics[width=0.48\textwidth,trim = 0mm 0mm 0mm 0mm,clip]{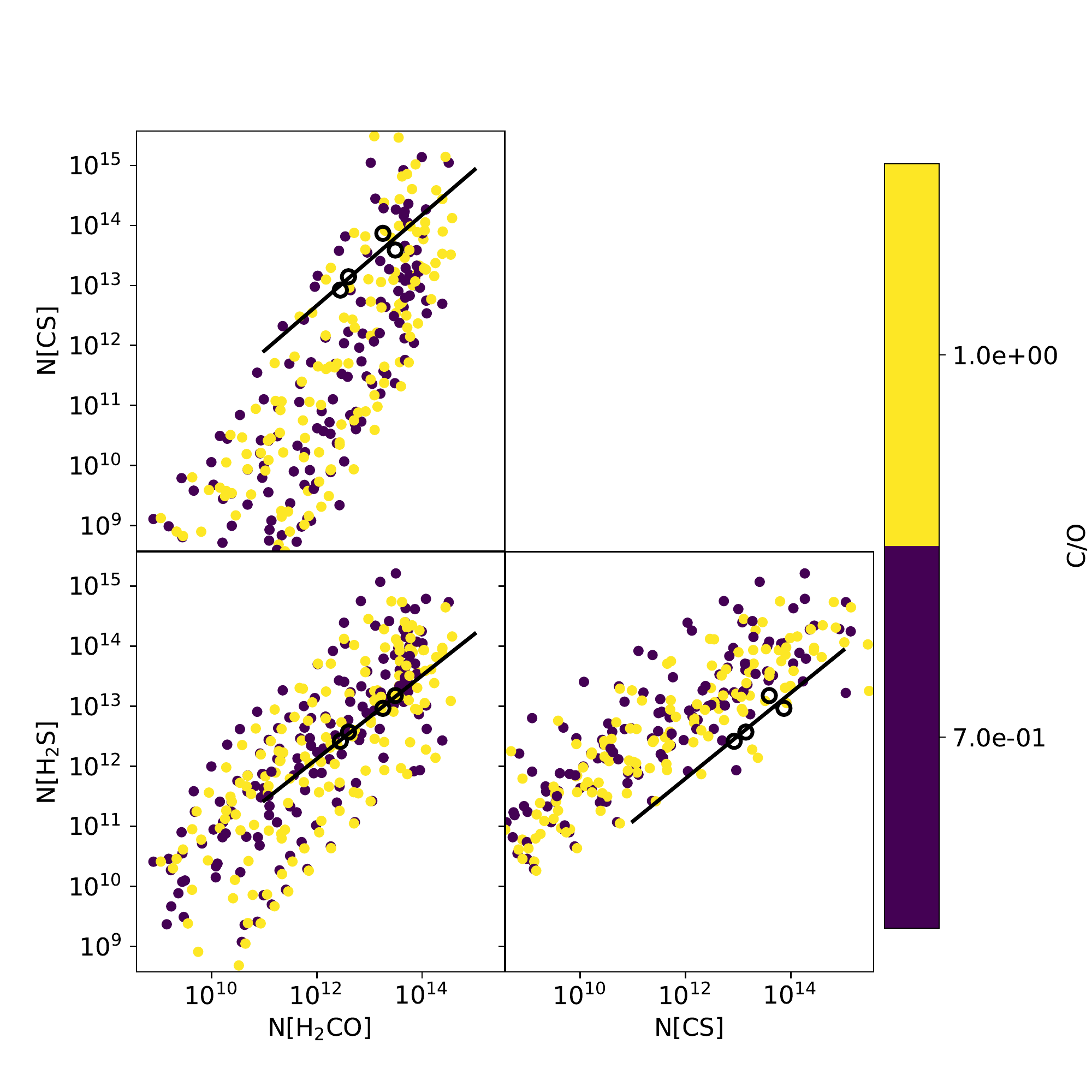}
  \caption{ Comparison of the observed correlations with the \texttt{Nautilus} grid of models using two different color codes: $\rm T_g$ (top left), $\rm n_H$ (top right), $\rm [S/H]$ (bottom left), and C/O (bottom right).}
 \label{Fig:model_grid}
\end{center}
\end{figure*}

We show in Fig. \ref{Fig:model_grid} the comparison of our observations with the grid of models. Since \texttt{Nautilus} does not perform radiative transfer, what we compare is the column density from the models with the column densities derived in Sect. \ref{Sect:col_dens}. Since the column densities are derived to fit the observed peak intensities, comparing line intensities is equivalent to comparing column densities. And since the derived column densities depend on the assumed source size, the exact position of the observations in these diagrams is subject to uncertainty. However, the shape of the correlation (slope and intercept) will be the same.  As can be seen, the models reproduce the correlations, although the slopes and intercepts might differ. Our aim is not to produce a perfect match to the observations, but rather to reproduce the correlation.

The \ce{H2S} versus \ce{H2CO} and \ce{H2S} versus \ce{CS} models and observations show very similar slopes, while in the case of \ce{CS} versus \ce{H2CO} our data points show a slope that is hard to match with the models. The fact that we observe correlations (\ce{H2S} vs. \ce{H2CO} and \ce{H2S} vs .\ce{CS}) in our small sample arises from the fact that these lines correlate for a large range of physical parameters ($\rm T_g$ and $\rm n_H$). In the case of \ce{H2S} versus \ce{H2CO,} the models point to temperatures ranging between 25 K and 45 K and densities between 10$\rm ^{6}~cm^{-3}$ and 3$\rm \times 10^{7}~cm^{-3}$. \ce{The case of H2S} versus \ce{CS} points to lower temperatures, between 20~K and 40~K, and low densities, between 10$\rm ^{4}~cm^{-3}$ and $\rm \times 10^{6}~cm^{-3}$. It is interesting to note that the \ce{H2S} versus \ce{H2CO} correlation is the one that lasts for a larger range of temperatures and densities. We also note that  \ce{H2S} versus \ce{H2CO} and \ce{H2S} versus \ce{CS} match better with sulfur-depleted models, while \ce{CS} versus \ce{H2CO} is better reproduced by models with solar sulfur abundance. The fact that we do not reproduce the \ce{CS} versus \ce{H2CO} slope and the need for solar sulfur abundance can both be explained by a lack of proper knowledge regarding the formation and destruction routes for CS. Several theoretical and observational works have pointed out the difficulty in simultaneously explaining the abundances of H$_2$S and CS using a single value of the elemental sulfur abundances and using the current gas and surface chemical networks \citep{Navarro2020, Bulut2021}. Furthermore, CS is an abundant species in different interstellar environments, such as discs, envelopes, and shocks \citep{Snell1984, Sakai2016, Oya2019, Taquet2020}. Minor contributions from different mechanisms might explain this divergence. Regarding the C/O ratio, the comparison of models and observations does not favor either of the two assumed values.

\section{Discussion \label{Sect:discussion}}
Our detection of o-\ce{H2S} in four protoplanetary systems in Taurus comes after its detection in GG Tau by \cite{Phuong2018}. Assuming LTE, $\rm n_{H_{2}}=10^7~cm^{-3}$, $\rm T_k = 20 K$,  and a disk size r = 500 au, we derived column densities in the range $\rm 2.6\times10^{12}~cm^{-2}$  to $\rm 1.5\times 10^{13}~cm^{-2}$. This is one order of magnitude larger than the value computed by \citet{Phuong2018} using DiskFit  toward GG Tau. In Fig. \ref{Fig:survey_col_dens} we show the CS, H$_2$CO, and H$_2$S column densities derived for the disks in our sample together with GG Tau. For GG Tau, we adopted the N(\ce{H2S}) value from \citet{Phuong2018}, while those for \ce{CS} and \ce{H2CO} were computed using RADEX to fit the line fluxes provided in \citet{Guilloteau2013} and \citet{Guilloteau2013}. The column densities of all molecules decrease from GV Tau to GG Tau, likely following an evolutionary sequence in which the molecular gas is progressively dispersed from the Class 0 to the Class III stages (only Class I and II sources are present in our sample). The GG Tau \ce{H2S} column density lies in the region below $\rm 10^{13}~cm^{-2}$, where we provide upper limits to H$_2$S column densities for AA Tau, DL Tau, and RY Tau. Since there seems to be a hierarchical relation between the column densities, and since our N(CS) upper limits are below the column density for GG Tau, we would expect N(\ce{H2S})$\rm < 10^{12}~cm^{-2}$ for these sources. We note that our estimates of the column densities for CS, \ce{H2CO}, and \ce{H2S} are based on a single transition and are therefore subject to large uncertainties. However, an idea of how large these uncertainties are can be obtained by comparing the column densities derived assuming different temperatures. In Sect. \ref{Sect:col_dens} we showed that by changing the temperature from 20 K to 30 K or to 10 K the column density of \ce{H2S} changed by $\rm \sim 8\%$ and $\rm \sim 25\%$ respectively, and therefore we have indications that our estimates of N(\ce{H2S}) are reliable. It is also interesting to compare the results around these young disks  with those in dark clouds.  \citet{Navarro2020} carried out a detailed study of the H$_2$S chemistry toward the dark cores TMC1-C and Barnard 1b. They found that the \ce{H2S} abundance is enhanced in the outer part of the envelope, where photodesorption and chemical desorption are most efficient. Our \ce{H2S} column densities in GV Tau, T Tau, HL Tau, and UY Aur compare well with the values found in the TMC 1 molecular cloud, suggesting that the emission could originate from the surfaces of the cold disk. As mentioned below, this is consistent with our chemical model.

When investigating the chemical evolution of disks, it is interesting to explore the column density ratios that are not affected by differences \textbf{ fluctuations?}in the total amount of molecular gas throughout the evolution of young stellar objects (YSOs). The column density ratios given in this paragraph were derived from column densities computed assuming beam dilution (see Sect. \ref{Sect:col_dens}). The sources with a larger N(\ce{H2S}), namely GV Tau and T Tau, show N(\ce{o-H2S})/N(\ce{o-H2CO}) ratios $\sim$0.5, while those with smaller N(\ce{H2S}) show a larger  N(\ce{o-H2S})/N(\ce{o-H2CO}) ratio of $\sim$0.9. If this is an evolutionary trend, it implies that \ce{H2CO} is more efficiently depleted onto dust grains compared to H2S in evolved disks. The N(\ce{H2S})/N(CS) ratio shows values ranging from 0.12 to 0.38. Carbon sulfide is an ubiquitous species that is very abundant in protostellar envelopes \citep{Sakai2014, Sakai2016} and molecular outflows \citep{WolfChase1998,Nilsson2000, Zhang2000,Li2015}, and it is not surprising that its column density sharply decreases as the protostellar envelope disappears and the molecular outflow declines. The outflow-envelope component seems less important in \ce{o-H2CO} and \ce{o-H2S}. We note that UY Aur is a Class II star, where we do not expect an envelope to be present. To validate the observed evolutionary trends, as well as the tentative correlations, we need a larger sample that allows us to compute reliable statistics .

We identified a possible correlation between the \ce{H2S} and \ce{H2CO} lines fluxes that can be explained by the fact that the formation of \ce{H2S} and \ce{H2CO} is dominated by surface chemistry reactions, while CS is preferentially formed via gas-phase reactions. In the NOrthern Extended Millimeter Array (NOEMA) observations of GG Tau by \cite{Phuong2018}, \ce{H2S} emission was observed from r $\sim$ 0.5$\arcsec$ to r $\rm \sim$ 2.5\arcsec, or 100 to 375 au. It is worth noting that all the stars where we detect \ce{H2S} belong to multiple systems, and so the role of multiplicity in the chemistry of young protoplanetary systems should be explored. The comparison of the CS and \ce{H2S} column densities with models seems to favor a large sulfur depletion ($\rm S/H=8\times10^{-8}$).We note, however, that the sulfur-depleted model overestimates \ce{H2CO}. N(\ce{H2CO}) is sensitive to other parameters, such as the initial conditions and the C/O ratio. A detailed exploration of the parameter space is needed in order to understand the chemistry in these systems, to conclude whether there is a large sulfur depletion onto dust grains or not, and to solve this discrepancy.

The sources in the sample were selected because they have been detected through water emission. Eight of them showed water emission at 63 $\rm \mu m$ \citep{Riviere2012}. We have identified a tentative correlation between the \ce{H2O} line emission at 63 $\rm \mu m$ and millimetric \ce{H2S} emission. According to \citet{Riviere2012}, the water emission originated in the inner regions of the disk, from 0.6  to 3 au.  If the correlation is real, it would imply that the properties of the inner and the outer disk might be coupled at the early stages of disk evolution.

\section{Summary and conclusions \label{Sect:summary}}
Surveying sulfuretted species in protoplanetary disks is key to understanding the evolution of sulfur chemistry during the planet formation stage. In the present paper we show the results of a single-dish survey of the CS 3-2 ($\rm f_0 = 146.969~ GHz$), \ce{H2CO} $\rm 2_{1,1}-1_{1,0}$ ($\rm f_0 = 150.498~ GHz$), and \ce{H2S} $\rm 1_{1,0}-1_{0,1}$ ($\rm f_0 = 168.763~ GHz$) lines carried out with the IRAM 30 m telescope. Our main results are the following.\\

\noindent
1. We detected  CS 3-2 and \ce{H2CO} $\rm 2_{1,1}-1_{1,0}$ lines toward six out of nine observed sources and \ce{H2S} $\rm 1_{1,0}-1_{0,1}$ lines toward four of them, namely GV Tau, HL Tau, T Tau, and UY Aur. This adds to the previous detection toward GG Tau, thus increasing the number of \ce{H2S} detections toward protoplanetary disks from only one to five. \\

\noindent
2. We have identified tentative correlations between \ce{H2S} and \ce{H2CO} line emission, as well as between \ce{H2S} and \ce{H2O}. This could be explained by a common origin of these species on dust grains.\\

\noindent
3. Assuming T=20 K, a gas density of 10$\rm ^{7}~cm^{-3}$, and using RADEX, we derived \ce{H2S} column densities in the range 2.6$\rm \times 10^{12}~cm^{-2}$ to $\rm 1.5 \times 10 ^{13}~cm^{-2}$.\\

\noindent
4. We used the astrochemical code \texttt{Nautilus} to build a grid of models that can be compared with our observations by varying the gas temperature and density. The grid of models reproduces the observed correlations for a large range of physical conditions.\\

\noindent 
5. GV Tau shows a particularly low  N(\ce{H2S})/N(CS) ratio when compared with the rest of the sample, a fact that can be interpreted as an evolutionary effect.\\

 Single-dish observations over a larger sample are needed to confirm our tentative correlations and evolutionary trends. Furthermore, interferometric observations are required to study the spatial distribution of the \ce{H2S} emission and validate our astrochemical model. Nevertheless, we proposse \ce{H2S} as a prominent reservoir of sulphur in protoplanetary disks.

\begin{acknowledgements}
We thank the Spanish MINECO for funding support from AYA2016-75066-C2-1/2-P and PID2019-106235GB-I100. IM is funded by a ``Talento'' Fellowship (2016-T1/TIC-1890, Comunidad de Madrid, Spain).  BM is partially funded by the Spanish Ministerio de Ciencia, Innovaci\'on y Universidades" through the national project ``On the Rocks II" (PGC2018-101950-B-100; PI E. Villaver).
\end{acknowledgements}

\bibliographystyle{aa} 
\bibliography{biblio}

\end{document}